\RequirePackage[2020-02-02]{latexrelease}
\documentclass[aps,pre, twocolumn, nofootinbib, showpacs,superscriptaddress,groupedaddress]{revtex4}  % for double-spaced preprint
\expandafter\let\csname equation*\endcsname\relax
\expandafter\let\csname endequation*\endcsname\relax

\usepackage{graphicx}  % needed for figures
\usepackage{amsfonts, amssymb}   % for math
\usepackage{bm}        % for math
\usepackage{amsmath}    % need for subequations
\usepackage{verbatim}   % useful for program listings
\usepackage{color}      % use if color is used in text
\usepackage{subfigure}  % use for side-by-side figures
\usepackage[colorlinks=true, citecolor=magenta, linkcolor=blue, filecolor=magenta, urlcolor=blue]{hyperref}   % use for hypertext links, including those to external documents and URLs
\usepackage{float}
\usepackage{physics}
\usepackage{enumerate}
\usepackage{mathrsfs}
\usepackage{relsize}
\usepackage{mathtools}
\usepackage{tikz}
\usetikzlibrary{graphs, arrows,chains,matrix,positioning,scopes,shadows}
\usepackage{natbib}

\newcommand{\half}{\frac{1}{2}}

\newcommand{\kae}[1]{ \left|#1 \right)}
\newcommand{\bae}[1]{\left(#1\right| }
\newcommand{\inprc}[2]{\left(#1\mid#2\right)}

\newcommand{\seaexp}[3]{\left(#1\mid#2\mid#3\right)}
\newcommand{\set}[1]{\mathscr{#1}}
\begin{document}

\title{Steepest Entropy Ascent Solution for a Continuous-Time Quantum Walker}
\author{Rohit Kishan Ray}
\email{rohitkray@iitkgp.ac.in}
\affiliation{Department of Physics, Indian Institute of Technology Kharagpur, India - 721302}
\date{\today}

%\begin{document}

\begin{abstract}
	We consider the steepest entropy ascent (SEA) ansatz to describe the non-linear thermodynamic evolution of a quantum system. Recently this principle has been dubbed the fourth law of thermodynamics  (Beretta Gian Paolo. 2020 The fourth law of thermodynamics: steepest entropy ascent. \href{https://doi.org/10.1098/rsta.2019.0168}{\textit{Phil. Trans. R. Soc. A.} \textbf{378:} 20190168.}). A unique global equilibrium state exists in this context, and any other state is driven by the maximum entropy generation principle towards this equilibrium. We study the SEA evolution of a continuous-time quantum walker (CTQW) on a cycle graph with N nodes. SEA solutions are difficult to find analytically. We provide an approximate scheme to find a general single-particle evolution equation governed by the SEA principle, whose solution produces dissipation dynamics. We call this scheme fixed Lagrange's multiplier (FLM) method. In the Bloch sphere representation, we find trajectories traced out by the Bloch vector within the sphere itself. We have discussed these trajectories under various initial conditions for the case of a qubit. A similar dissipative motion is also observed in the case of CTQW, where probability amplitudes have been used to characterize decoherence. Our FLM scheme shows good agreement with numerical results. As reported in the text, in CTQW, a strong delocalization exists for low system relaxation time.
\end{abstract}

\maketitle
\section{Introduction}
In the era of quantum technologies, quantum algorithms have occupied an important role. Due to the discovery of better-than-classical performance of Shor's algorithm \cite{ShorPolyTime}, and Grover's search algorithm \cite{Grover1}, followed by a plethora of findings, it has become an active area of research to find such algorithms or improve upon existing ones. One such candidate is the quantum analog of the classical random walk, \textit{the quantum walk} \cite{Aharonov1, Childs1, Shenvi1, Farhi2, Manouchehri1, childs2009universal, fourier1, fourier2, jpd1, jpd2}. Quantum walks are of two kinds, discrete-time quantum walk (DTQW) and continuous-time quantum walk (CTQW) (for a review, see Refs. \cite{Manouchehri2014,kendon_2007, Fedichkin2021}, and references therein). Despite having similar probability distributions and features, discrete-time and continuous-time kinds are not equivalent; although, under certain limits, DTQW can produce CTQW \cite{Childs2010, DALESSANDRO201085, StrauchContDisc}. In this paper, we focus on CTQW only. Their quadratic speed-up over classical walks characterizes quantum walks (QW) \cite{kendon_2007}. QWs have been performed on graphs \cite{aharonov2001quantum}. Quantum walks can be used to perform a spatial search \cite{Childs1, Shenvi1}. A scheme for quantum computation has been provided in Ref. \cite{childs2009universal}, where the universality of QW was explored. Many body QWs have been studied to understand the nature of entanglement in multipartite systems \cite{fourier2, jpd1,jpd2, rohde1,rohde2}.

Besides studying unitary quantum evolution, many scholars have studied decoherence and dissipation in quantum walks \cite{fourier1, roma3, roma2, roma1, kendonDecoherence2003, fedichkin2005mixing}. Introduction of multi-coin walks leads to dissipation \cite{fourier1}. Others have considered decoherence introduced by measurements on the vertices of a graph \cite{kendonDecoherence2003}. Alternatively, some have considered decoherence on the particle, coin, or both \cite{fedichkin2005mixing}. \citet{kendonDecoherence2003} have shown decoherence to be a useful facet in quantum evolution. As the walker is an otherwise isolated system, decoherence is introduced as the action of certain measurements on the system in an operational setting \cite{nielsen}. Understanding mixing time under such modeling is an active area of research \cite{shantanavMixing2020,shantanavMarkovMixing2020}. On the other hand, by using the concept of quantum stochastic walks \cite{whitefieldQSW}, decoherence in CTQWs has been explored by \citet{silva1} using Lindbald-type master equations. \citet{roma3} split the evolution into two parts, one contributing to the Markovian process, while the other part to the interference responsible for unitary evolution. A novel way of studying thermodynamics has been introduced and followed in Refs. \cite{roma2,roma1} by varying the contribution due to the interference factor. \citet{quadraticwalk} have used quadratic perturbation in the Hamiltonian to study CTQW. However, we are interested in a more fundamental aspect of decoherence - as a first-principle result rather than one of a phenomenological origin. Such a theory should reproduce quantum results under proper limits as desired, besides attempting to explain nonequilibrium behavior.

Studying an open quantum system involves introducing non-unitary behavior in the dynamics. It can be achieved by introducing some super-operators with anti-commutation relations invoking dissipation in the system, resulting in a phenomenological origin of the entropy generation process \cite{li2016von}. Consequently, a theory of thermodynamic evolution of quantum states through modeling at different \textit{levels of description} to arrive at the equation of motion has been in vogue \cite{grmelaGENERIC1, grmelaGENERIC2}. Theories belonging to this category can be classified as a general equation for the nonequilibrium reversible-irreversible coupling (GENERIC). In this setup, microscopic details of the fundamental structure are ignored in favor of macroscopic dynamics. As a consequence, the constraints become global invariants of motion. The GENERIC theory essentially lacks a local description, which can be modeled using a metric-dependent relaxation parameter \cite{Beretta4}, and a concept of equilibrium to drive the system. Let us consider global and local equilibriums in the Lyapunov sense \cite{berettalyapunov}. We find an evolution-trajectory in the state space characterized by maximization of the entropy production rate under the constraint of invariants of the motion \cite{Hatsopoulos76_I, Hatsopoulos76_IIa, Hatsopoulos76_IIb, Hatsopoulos76_III,Beretta1984_1, Beretta1985, Beretta1985_2, Beretta_SEA_87, gheorghiu1,gheorghiu1a,beretta12}. In addition to this, consider the second law of thermodynamics: the existence of a unique globally stable state to satisfy the second law requirement \cite{Beretta2020} (A rich review of literature in this domain is compiled in the Ref. \cite{nonequilibriumthermodynamics}). The unitary trajectories described by Schr\"{o}dinger type evolution form limit cycles in this notion \cite{beretta1}. As a result, states that belong to the quantum mechanical evolution are extrema of the set containing all such states considered in the description. This description also requires the identification of the density matrix as a valid state operator and $\rho^2 = \rho$ to be the defining feature of pure QM limit cycles. In the case of a qubit, the rest of the $\trace(\rho^2)\le 1$ states thus fill up the entire Bloch-sphere and represent the state space available for non-linear evolution.

The steepest entropy ascent (SEA) dynamics framework finds its roots in the first-principle approach to understand \textquotedblleft spontaneous decoherence\textquotedblright. Moreover, the SEA formalism can be extended to other disciplines of studying non-linear evolution of a more general kind \cite{beretta2014steepest}. In follow-up work, under similar kinematic considerations, the mathematical equivalence between the non-linear contribution due to GENERIC and SEA has been drawn by \citet{montefusco2015essential}. While working in a far-off equilibrium situation, defining temperature becomes troublesome, as the temperature is defined using steady states. The concept of hypoequilibrium as invoked in Refs. \cite{liVonSEA2016,kim1} seems to serve this purpose. In this SEA theory, we wish to solve for the continuous-time quantum walker and analyze the solution.

In a nutshell, quantum walks can be useful with decoherence. This decoherence is usually modeled through measurements performed at each walk instance, and the walker state mixes according to some rule. However, when a first-principle approach (SEA type dynamics) involving a local description and an entropy non-decrease postulate is considered to study the relaxation of a system subject to initial perturbation, decoherence results without external interaction. We consider a single quantum walker walking on a cycle of $N$ nodes in continuous time, whose evolution is guided by the SEA formalism. Its solution can be applied to any time-independent Hamiltonian containing a single particle. We show that the time-dependent relaxation dynamics characterize this evolution. Under such a scenario, both SEA and unitary dynamics can be recovered. Section \ref{prelims} introduces the readers to preliminary physical ideas. Here we discuss the basics of the continuous-time quantum walk and the building blocks of the SEA principle to keep the paper self-contained. In section \ref{Solution}, we first derive an approximate analytical solution. As the SEA equation of motion (EoM) (equation (\ref{req}) below) is highly non-linear, it is difficult to solve analytically \cite{Beretta1985,gheorghiu1}, and we resort to numerical solutions. This limits our ability to understand the dynamics analytically. Interestingly though, as shown below in section \ref{resulta}, considering a fixed Lagrange's multiplier (FLM) approach eases the inherent non-linearity of the EoM, making the solution analytically tractable. As an application of this, we solve for a two-level system: qubit, which closely agrees with the exact result \cite{Beretta1985} present in the literature. In the second part of section \ref{Solution}, we solve for the $N$ level system- a quantum walker and show plots of the solutions followed by our analysis. We have delegated detailed derivations to the appendices to preserve the flow of this paper. In some cases, the readers are requested to go through the references provided. In section \ref{conclusion}, we discuss the importance of our result and conclude by pondering further questions that have come up through this work. The appendices are structured as follows- in Appendix \ref{Appendix1}, we have provided the geometric interpretation of SEA evolution. We have shown how the SEA EoM is derived using the variational principle. In the following Appendix \ref{Appendix2}, we derive the expressions for Lagrange multipliers and provide the complete SEA equation of motion. Appendix \ref{Appendix3} contains comments on relaxation time, and in Appendix \ref{Appendix4}, we have computed the Lagrange multipliers for the CTQW.

%%%%%%%%%%%%%%%%%%%%%%%%%%%%%%%%%%%%%%%%%%%%%%%%%%%%%%%%%%%%%%%%%%%%%%%%%%%%%%%%%%%%%%%%%%%%%%%%%%%%%%%%%%%

\section{\label{prelims}Theoretical Preliminaries}

Consider a quantum walker walking on some undirected graph $ \mathcal{G} $ with $N$ vertices, vertex, and edge set as $\mathbb{V}$, and $\mathbb{E}$ respectively. $\mathcal{G}$ has no double edge or self-loops. The adjacency matrix $\mathbf{A}$ of $\mathcal{G}$ can be defined as follows,
\begin{equation*}
	\mathbf{A}: a_{ij} = \begin{cases}
		1 & \text{if } e_{ij} \in \mathbb{E} \\
		0 & \text{otherwise}.
	\end{cases}
\end{equation*}
Thereafter, we can define the Laplacian $\mathbf{L}$ of $\mathcal{G}$ as \cite{Childs1, farhi1},
\begin{equation*}
	\mathbf{L} = \mathbf{D}-\mathbf{A},
\end{equation*}
$\mathbf{D}$ is diagonal and has an entry as the degree of the $i^{\text{th}}$ vertex, $d_i$. We associate a hopping probability $\mu_{ij}$ with the probability of transition between two adjacent vertices ($v_i$, $v_j$ ) per unit time. Considering uniform transition rates $\mu_{ij}=\mu$, for the unitary continuous-time quantum walker we can write \cite{Manouchehri2014,kendonDecoherence2003,Childs1},
\begin{equation}\label{qweom}
	\derivative{\ket{\Psi}}{t} = -\frac{i}{\hbar}\mu\mathbf{L} \ket{\Psi}.
\end{equation}
In the equation (\ref{qweom}), the quantity $\mu \mathbf{L}$ is identified as the Hamiltonian $ \mathcal{H} $ of the CTQW. The solution to equation (\ref{qweom}) is read as,
\begin{equation}\label{qwsol}
	\ket{\Psi(t)} = \exp(-i\mu \mathbf{L} t)\ket{\Psi(0)}\equiv \mathcal{U}(t)\ket{\Psi(0)}\equiv\mathcal{U}_t\ket{\Psi(0)},
\end{equation}
where, $\ket{\Psi(0)}$ is the initial state of the walk, and $\hbar=1$. In terms of density matrix $\rho$,\footnote{$t$ is added to superscript as label for a time, instead of using the standard $\rho(t)$, to make the notations less cumbersome.} we can write equation (\ref{qwsol}) as,
\begin{equation}\label{qwrho}
	\rho^t= \mathcal{U}_t\rho^0\mathcal{U}_t^{\dagger},
\end{equation}
implying that the quantum state of the walker undergoes unitary rotation in the state space as the walker exhibits CTQW.

Steepest entropy ascent motion talks about the relaxation of a system away from equilibrium. The interested readers are directed towards Refs. \cite{beretta2014steepest,Beretta2020,Beretta5,liVonSEA2016} for a detailed understanding of SEA dynamics. Here, we explore the observed results while delegating a basic introduction to Appendix \ref{Appendix1}, where the derivation of the SEA EoM is done in brief. In the SEA evolution, the dynamical equation can be written in the general Ginzburg-Landau form \cite{liVonSEA2016} as under,
\begin{equation}\label{eq1}
	\derivative{\rho}{t} = -i\comm{\mathcal{H}}{\rho}+\frac{1}{\tau}\acomm{\mathcal{D}}{\rho}.
\end{equation}
Where $\acomm{\cdot}{\cdot}$ represents anti-commutator, $\tau$ relaxation time, and $\mathcal{D}$ is related to dissipation, the non-unitary part of the evolution. In the SEA formalism, the state operator $\gamma \equiv \kae{\gamma}$ is an element of linear manifold $\set{L}$, a state-space in Hilbert space $\set{H}$, with a symmetric inner product $\inprc{A}{B}=  \Tr(A^{\dagger}B+B^{\dagger}A)/2$ \cite{beretta2014steepest}. To maintain a positive $\rho$ for all times of SEA motion, the squared root of $\rho$ is considered the state operator and is computed using spectral theorem \cite{Beretta4}, $\rho = \gamma\gamma^{\dagger}$. Let the set $\{\mathbf{C_i}(\gamma)\}$ contain operators denoting various conservation quantities to be dealt as constraints of motion, such as energy ($\mathcal{H}$), number of particles ($\mathcal{N}$), etc. The functional derivative of $\mathbf{C_i}$ is denoted by $\kae{\Psi_i} = \kae{\delta \mathbf{C_i}(\gamma)/\delta\gamma}$. Similarly, the entropy functional is ($k\equiv k_B$, Boltzmann's constant),
$$ \mathbf{S} = -k \Tr(\rho \ln(\rho)),$$
and its functional derivative is $\kae{\Phi}=\kae{\delta \mathbf{S}(\gamma)/\delta\gamma}$ \cite{beretta2014steepest,Beretta5}. The constraint of entropy non-decrease and conservation of $\mathbf{C_i}$ can be expressed as,
\begin{eqnarray}
	\derivative{\mathbf{S}}{t} = \Pi_S & \quad \text{with} \quad & \Pi_S = \inprc{\Phi}{\Pi_{\gamma}} \ge 0, \label{req3} \\
	\derivative{\mathbf{C_i}}{t} = \Pi_{C_i} & \quad \text{with} \quad & \Pi_{C_i} = \inprc{\Psi_i}{\Pi_{\gamma}} = 0. \label{req4}
\end{eqnarray}
Using the above constraint equation, and by applying Lagrange's multiplier method, the SEA rate of change of state operator, $\derivative{\gamma}{t} = \Pi_{\gamma}$ is given below,
\begin{equation}\label{reqm}
	\kae{\Pi_{\gamma}} = \mathcal{L}\kae{\Phi - \sum_{i}\beta_i\Psi_i}.
\end{equation}
Where the metric associated with the manifold $\set{L}$, $\hat{G}(\gamma)$ appears as $\mathcal{L} = \frac{1}{\tau}\hat{G}(\gamma)^{-1}$ \cite{beretta2014steepest}. This can be re-written in terms of $\rho$ (see Appendix \ref{Appendix1} for details) for an uniform (Fisher) metric as,
\begin{equation}\label{req}
	\dv{\rho}{t} = -\dfrac{1}{\tau}\left[\rho\ln(\rho)+\half\sum_i(-1)^i\beta_i\acomm{\mathbf{C_i}}{\rho}\right] - i\comm{\mathcal{H}}{\rho}.
\end{equation}
On comparing with equation (\ref{eq1}), we can write $\acomm{\mathcal{D}}{\rho} = -\left[k\rho\ln(\rho)+\half\sum_i(-1)^i\beta_i\acomm{\mathbf{C_i}}{\rho}\right]$. Choosing $\tau, \beta_i$ as Lagrange's multipliers, the full SEA evolution, including the expressions for the $ \beta_i $'s, can be written in the following compact form conveying all the necessary information (see Appendix \ref{Appendix2}).
\resizebox*{.8\linewidth}{!}{
	\begin{minipage}{\linewidth}
		\begin{align}\label{seacompact}
			\begin{split}
				&\dv{\rho}{t} +i\comm{\mathcal{H}}{\rho} =\\
				&  -\dfrac{1}{\tau}\tfrac{\vmqty{\rho\ln(\rho) & \half\acomm{\mathbf{C_1}}{\rho} &  \half\acomm{\mathbf{C_2}}{\rho} &  \half \acomm{\mathbf{C_3}}{\rho}\\
						\tr(\tfrac{\rho}{2}\acomm{\mathbf{C_1}}{\ln(\rho)}) & \tr(\rho\mathbf{C_1}^2) &  \tr(\tfrac{\rho}{2}\acomm{\mathbf{C_1}}{\mathbf{C_2}}) & \tr(\tfrac{\rho}{2}\acomm{\mathbf{C_1}}{\mathbf{C_3}})\\ \tr(\tfrac{\rho}{2}\acomm{\mathbf{C_2}}{\ln(\rho)}) &\tr(\tfrac{\rho}{2}\acomm{\mathbf{C_2}}{\mathbf{C_1}}) & \tr(\rho\mathbf{C_2}^2) & \tr(\tfrac{\rho}{2}\acomm{\mathbf{C_1}}{\mathbf{C_3}})\\
						\tr(\tfrac{\rho}{2}\acomm{\mathbf{C_3}}{\ln(\rho)}) &	 \tr(\tfrac{\rho}{2}\acomm{\mathbf{C_3}}{\mathbf{C_1}}) & \tr(\tfrac{\rho}{2}\acomm{\mathbf{C_3}}{\mathbf{C_2}}) & \tr(\rho\mathbf{C_3}^2)}}{\vmqty{\tr(\tfrac{\rho}{2}\acomm{\mathbf{C_1}}{\mathbf{C_1}}) &  \tr(\tfrac{\rho}{2}\acomm{\mathbf{C_1}}{\mathbf{C_2}}) & \tr(\tfrac{\rho}{2}\acomm{\mathbf{C_1}}{\mathbf{C_3}})\\
						\tr(\tfrac{\rho}{2}\acomm{\mathbf{C_2}}{\mathbf{C_1}}) & \tr(\tfrac{\rho}{2}\acomm{\mathbf{C_2}}{\mathbf{C_2}}) &\tr(\tfrac{\rho}{2}\acomm{\mathbf{C_2}}{\mathbf{C_3}})\\
						\tr(\tfrac{\rho}{2}\acomm{\mathbf{C_3}}{\mathbf{C_1}}) & \tr(\tfrac{\rho}{2}\acomm{\mathbf{C_3}}{\mathbf{C_2}}) & \tr(\tfrac{\rho}{2}\acomm{\mathbf{C_3}}{\mathbf{C_3}})}}.
			\end{split}
		\end{align}
	\end{minipage}}

To solve for CTQW, we identify $\mathbf{C_2}$ with $\mu \mathbf{L}$, and use probability conservation ($ \mathbf{C_1} = \mathbf{I} $). We do not need $ \mathbf{C_3} $ for a single walker; the number operator becomes redundant. Thence,
\begin{equation}
	\mathcal{D} = -\left[k\ln(\rho)-\beta_I\mathbf{I}+\beta_H\mu\mathbf{L}\right].
\end{equation}
Substituting this expression for $\mathcal{D}$ in equation (\ref{eq1}), we solve for $\rho$. Interestingly, we note that equation (\ref{req}) is free of $\gamma$, making it a deterministic process\cite{beretta1}.

%%%%%%%%%%%%%%%%%%%%%%%%%%%%%%%%%%%%%%%%%%%%%%%%%%%%%%%%%%%%%%%%%%%%%%%%%%%%%%%%%%%%%%%%%%%%%%%%%%%%%%%%%%%%%%%%%%%%%%%%%%%%
\section{\label{Solution}Approximate analytical solutions of the full SEA evolution}
\subsection{\label{resulta}Single particle case (arbitrary dimension)}
In this section, we intend to solve equation (\ref{req}) for a single particle in arbitrary dimensions under a given Hamiltonian $\mathcal{H}$. The motion is constrained by the condition of entropy $\mathbf{S}$ non-decrease, along with conservation of quantities $ \mathbf{C_i} $ such as energy, probability, etc. These conditions are introduced via Lagrange multipliers $\beta_i  $s in equation (\ref{req}). These $\beta_i$ are non-trivially dependent on $\rho$, which makes it difficult to solve equation (\ref{seacompact}) analytically with such non-linearity; therefore, we resort to numerical techniques. However, an approximate analytical solution can help us understand the essence of the general non-linear evolution.

In this work, we introduce the following approximation. We consider the $\beta_i$'s (defined in equation (\ref{seaqa2})) to be at their prefixed values, such as their initial or final values, and call this approximation fixed Lagrange's multiplier (FLM) method\footnote{ There is a natural motivation behind this assumption. Consider a real function $f(x,y)| x,y \in \mathbb{R}$, to be optimized under some constraint $g(x,y) =c$, where $c$ is a real constant. We can define a \textit{Lagrangian} function $L(x,y,c):= f(x,y)+\lambda(c-g(x,y))$ with $\lambda = \pdv{L}{c}$ as the Lagrange's multiplier. This suggests, for a fixed $c$ there exists a constant $\lambda$ such that $f(x,y)$ is optimized.}. FLM method enables us to have an approximate analytic expression for the single quantum constituent case, which produces results in good agreement with the numerical computation. We begin with a general $\rho$. Since $\rho$ is always positive, such a $\rho$ can have a diagonalization which can be achieved via a similarity transformation, a trace-preserving operation. Let us consider one such transformation on a given $\rho$ as follows,
\begin{equation*}
	\rho^t = \mathbf{U}_t \rho_d^t \mathbf{U}_t^{-1},
\end{equation*}
where $\mathbf{U}_t$ is formed by column-wise stacking the eigenvectors of $\rho$ at time $t$, and $\rho_d^t$ is the required diagonal matrix. Using this transformation in equation (\ref{req}) we get,
\begin{align}\label{eqdiag}
	\begin{split}
		\dv{\rho_d^t}{t} =& -\dfrac{1}{2\tau}\left[\acomm{\ln(\rho_d^t)}{\rho_d^t}+\sum_{i}(-1)^i\beta_i\acomm{\mathbf{C_i}^d}{\rho_d^t}\right]\\ & -i\comm{\mathcal{H}^d}{\rho_d^t}.
	\end{split}
\end{align}
Note that, $\mathcal{H}^d = \mathbf{U}_t^{-1}\mathcal{H}\mathbf{U}_t$, and $\mathbf{C_i}^d = \mathbf{U}_t^{-1}\mathbf{C_i}\mathbf{U}_t$ \footnote{Please take note that neither $\mathcal{H}^d$, nor $\mathbf{C_i}^d$ are the diagonal forms of the respective matrices (they could be, if they can be simultaneously diagonalized with $\rho$). I was running out of symbols; any confusion due to this is regretted.}.\\

Before we proceed, certain comments are necessary. First we note $\dv{\rho^t}{t} = \comm{\dot{\mathbf{U}_t}\mathbf{U}_t^{-1}}{\rho^t}+\mathbf{U}_t\dot{\rho}^t_d\mathbf{U}_t^{-1}$. Thus implying, in general, eigenvectors can have non-zero time-dependence. However, a restricted class of problems exists, where the eigenbasis of $\rho$ can change solely due to the Hamiltonian. In such a scenario, the commutator term appearing in the time derivative becomes zero, and we are left with constant $\mathbf{U}_t$. As a result, we have a special class of evolution where a $\rho$ initially diagonal in energy basis remains so throughout the dynamics. In the subsequent computations, we exploit this `diagonal property' of $\rho$ to find an analytic approximation. Although this limits the range of evolutions considerable under the given assumption, it still pertains to a broad class of solutions as supported by the examples presented in the paper.\\

$\rho$ has a spectral decomposition in its eigenbasis $\{\ket{\lambda_i}\}$, and $\rho_d$ is diagonal in the standard basis $\{\ket{i}\}$ ($\ket{i}$ is an $N$-dimensional unit vector) as follows,
\begin{align*}
	\rho_d = \sum_{ij}\lambda_i\delta_{ij}\dyad{i}{j}; & \quad\text{ and }
	\begin{split}
		\rho =& \sum_{i}\lambda_i\dyad{\lambda_i}{\lambda_i},\\
		=&\sum_{ij}\lambda_i\delta_{ij}\dyad{\lambda_i}{\lambda_j}.
	\end{split}
\end{align*}
$\delta_{ij}$ is the Kronecker delta. Using these expansions in equation (\ref{eqdiag}), and by choosing $\left[\mathcal{H}^d\right]_{ij}=H_{ij}^d$, and $\left[\mathbf{C_s}^d\right]_{ij}=C_{ij}^s$, the RHS of equation (\ref{eqdiag}) can be modified as,
\begin{align*}
	\begin{split}
		&-\frac{1}{2\tau}\sum_{ijm}\delta_{im} \Big[2\lambda_m\ln(\lambda_m)\delta_{mj}\\
		& +\sum_{s}(-1)^s\beta_s\left[\lambda_m+\lambda_j\right]C_{mj}^s\Big]\dyad{i}{j}\\ &-i\sum_{ij}\left[\lambda_j-\lambda_i\right]H_{ij}^d\dyad{i}{j}.
	\end{split}
\end{align*}

For a single particle, we are mostly interested in energy and probability conservation; the complete expression upon considering a Euclidean metric reads as
\begin{align*}
	\begin{split}
		\sum_{ij}\dot{\lambda}_i\delta_{ij}\dyad{i}{j} = & -i\sum_{ij}\left[\lambda_j-\lambda_i\right]H_{ij}^d\dyad{i}{j}\\
		&-\frac{1}{2\tau}\sum_{ij}\Big[2k\lambda_i\ln(\lambda_i)\delta_{ij}- 2\beta_I\lambda_i\delta_{ij}\\
		&+\beta_H\left[\lambda_j+\lambda_i\right]H_{ij}^d \Big]\dyad{i}{j}.
	\end{split}
\end{align*}
According to our assumption LHS of equation (\ref{eqdiag}) is diagonal,
\begin{equation}\label{eigenequation}
	\dv{\lambda_i}{t} = -\frac{1}{\tau}\left[\lambda_i\ln(\lambda_i)-\beta_I\lambda_i+\beta_H\lambda_i H^d_{ii}\right].
\end{equation}
For diagonal density matrices, similarity transformation is identity, so we get,
\begin{equation}\label{p_iequation}
	\dv{p_i}{t} = -\frac{1}{\tau}\left[ p_i\ln(p_i)-\beta_I p_i+\beta_H p_i H_{ii}\right],
\end{equation}
where $p_i=\left[\rho^t\right]_{ii}$. Both the equations (\ref{eigenequation}), and (\ref{p_iequation}) have a similar type of solution, namely that of almost identical non-linear ODE. Using standard techniques and FLM approximation, we arrive at the following expression,
\begin{equation}\label{p_isol}
	p_i(t) = \exp(\exp(w_i-\frac{t}{\tau})+\beta_I-\beta_H H_{ii}).
\end{equation}
We have $w_i=\ln(\ln(p^0_i)-\beta_I+\beta_H H_{ii})$, where $p^0_i$ is the $i^{\text{th}}$ diagonal entry of initial $\rho$. The solution produced above can be written in a straightforward form, identifying $\eta^c_i = \beta_H H_{ii}-\beta_I$, or for general cases as $\sum_{s}(-1)^s\beta_s C^s_{ii}$; $\tilde{v}_i = e^{w_i}$, and $\eta(t) \equiv \eta^t = \exp(-t/\tau)$, we get,
\begin{equation}\label{p_isimple}
	p_i(t) \equiv p^t_i = \exp(\tilde{v}_i\eta^t-\eta^c_i).
\end{equation}
And we find $\tilde{v}_i$ as,
\begin{align*}
	         & p_i(0) =  p_i^0,                                         \\
	         & \exp(\tilde{v}_i\eta^0-\eta^c_i) =  p_i^0,               \\
	\implies & \tilde{v}_i =                       \ln(p_i^0)+\eta^c_i.
\end{align*}
Hence, we can write $\rho_d^t = \sum_{i}p_i^t\dyad{i}{i}$. For a general initial $\rho$ with off-diagonal terms can be written as - $\rho = \mathbf{U}_t\rho_d^t\mathbf{U}_t^{-1}$, whereas if we consider only diagonal $\rho$'s, we get $\rho = \rho_d^t$. Including the Hamiltonian evolution, we get the following equation for uniform metric ($\mathcal{U}_t \equiv\exp(-i\mathcal{H}t)$, and projections $\mathbb{P}_m = \dyad{m}{m}$),
\begin{align}\label{fullevolution}
	\begin{split}
		\rho^t = & \\ &\mathcal{U}_t\mathbf{U}_t\left(\sum_{m}\exp(\eta^t_m-\eta^c_m)\mathbb{P}_m\right)\mathbf{U}_t^{-1}\mathcal{U}_t^{\dagger},
	\end{split}
\end{align}
where, $\eta^t_m = \left(\ln(p_m^0)+\eta^c_m\right)e^{-t/\tau} = \tilde{v}_i\eta^t$, and $\eta^c_m = \sum_{s}(-1)^s\beta_sC_{mm}^s$.
So far, as we can see, equation (\ref{fullevolution}) represents the evolution of $\rho^t$s diagonal in the energy basis, except that it only considers the Fisher metric. We use this equation to understand spontaneous decoherence in the evolution of a walker performing CTQW (see equations, (\ref{qweom}-\ref{qwrho})). We consider the case of a two-level system and use equation (\ref{fullevolution}) to observe its motion on the Bloch sphere. Despite the problem's simplicity, it is a sparsely explored area even within the community \cite{Beretta1985,Beretta2020}. A qubit evolution can help us understand some of the salient features of SEA dynamics. In this spirit, and for the sake of consistency in presentation, we explore the two-level system in detail to appreciate the results for the $N$-level system with relative ease.
%....................................................................................................................
\subsection{\label{qubit}Case I - Two-level system: Qubit}

A general two-level quantum system can be represented by a qubit with two possible outcomes of a measurement,\textit{i.e.}, level 1:  $\ket{0}$, and level 2: $\ket{1}$. The most general density matrix representing a qubit can be expressed in the following form,
\begin{equation*}
	\rho = \half\left(\mathrm{I}+\vec{r}\cdot\vec{\sigma}\right).
\end{equation*}
Where $\vec{\sigma}$ is the Pauli vector with components as Pauli matrices for a spin-$\half$ system, $\vec{r}$ is the radial vector of a Riemann sphere, also known as Bloch sphere \cite{nielsen}.

\begin{figure}[h]
	\centering
	\includegraphics[width=0.9\linewidth]{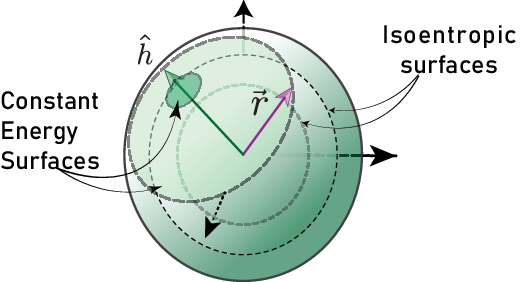}
	\caption{\label{bloch}(Color online) A Bloch sphere representation of a qubit. The purple arrow denotes the vector $\vec{r}$, while the green one $\hat{h}$. The isoentropic concentric surfaces and the constant energy planes are labeled in the diagram.}
\end{figure}

The most general Hamiltonian acting on a point on this sphere can be written as $\mathcal{H}=\left(\omega_0\mathrm{I}+\omega\hat{h}\cdot\vec{\sigma}\right)$, where $\hat{h}$ is a unit vector along the axis of rotation due to $\mathcal{H}$, and $\hbar=1$.  Eigenvalues of $\mathcal{H}$ are $h_{\pm}=\omega_0\pm\omega$, $h = \abs{\hat{h}} = 1$, and $2\omega$ is the precession frequency of the state vector around $\hat{h}$, $\omega_0$ is a constant $=\half\tr(\mathcal{H})$  \cite{Beretta1985}. FIG. \ref{bloch} shows that the states are constrained to lie on the intersection of constant energy and isoentropic surfaces. The figure shows that on each isoentropic surface, the radius vector traces out the intersection of constant energy planes (perpendicular to $\hat{h}$) with the surface.
We notice that to include mixed states in our description, $\vec{r}$ has variable magnitude $r^t$, and the maximally mixed state is represented by $\vec{r}=\vec{0}$. Considering $\tr(\rho^2) = R$, eigenvalues of $\rho$, $\lambda_{\pm}=\half\left(1\pm r\right)$, we can see immediately,
\begin{equation*}
	r=\sqrt{2R-1},
\end{equation*}
which further implies $0\le r \le 1$, for $\half\le R\le 1 $.

Using the eigenvalues $\lambda_{\pm}$ we can write the entropy function as:
\begin{align}\label{entropyonr}
	\begin{split}
		\mathbf{S}=& -k\left[\lambda_+\ln(\lambda_+)+\lambda_-\ln(\lambda_-)\right],\\
		=&-k\left[\frac{1+r}{2}\ln(\frac{1+r}{2})+\frac{1-r}{2}\ln(\frac{1-r}{2})\right],\\
		=& - \frac{k}{2}\ln(\frac{1-r^2}{4})-\frac{kr}{2}\ln(\frac{1+r}{1-r}).
	\end{split}
\end{align}
The isoentropic surfaces form concentric spheres, with entropy increasing as $r$ decreases. Therefore, the main dissipative part of the dynamics will try to draw any state within and on the surface of the Bloch sphere toward its center, where a maximally mixed state with maximum entropy $k\ln(2)$ resides. Since pure unitary dynamics is non-dissipative, it is required and customary to initially disturb (quench) a quantum mechanical state from the limit cycle of unitary evolution and then observe the system's relaxation as time progresses. The maximally mixed state is the global equilibrium ($\rho_u$) state for a completely isolated system. Thus, we quench the initial state ($\rho^0$) as follows \cite{beretta12},
\begin{align*}
	\begin{split}
		\rho = &\varepsilon \rho^0 + \left(1-\varepsilon\right)\rho_u\\
		\sum_{i}\lambda_i\dyad{\lambda_i}{\lambda_i}=& \sum_{i}\varepsilon\lambda_i^0\dyad{\lambda_i^0}{\lambda_i^0}+\left(1-\varepsilon\right)\frac{\mathrm{I}}{\tr(\mathrm{I})}.
	\end{split}
\end{align*}
If $\rho^0$ is diagonal, then
\begin{equation}\label{initial}
	\lambda'_i = \varepsilon\lambda_i^0+(1-\varepsilon)\frac{1}{N},
\end{equation}
where $N=\tr(\mathrm{I})$, which for this case is 2. $\varepsilon$ is a variable parameter $\in\left[0,1\right]$, with zero value denoting the completely mixed state. Armed with all these and a Euclidean metric, we consider equation (\ref{eigenequation}),
\begin{align*}
	\begin{split}
		\dv{\lambda_{\pm}}{t} = & -\frac{1}{\tau}\left[\lambda_{\pm}\ln(\lambda_{\pm})+(\beta_HH_{\pm}^d-\beta_I)\lambda_{\pm}\right]\\
		\implies\pm\dot{r} = & \mp\frac{1}{\tau}\Big[\left(1\pm r\right)\ln(\frac{1\pm r}{2}) \\
			& +(\beta_H H_{\pm}-\beta_I)\left(1\pm r\right)\Big].
	\end{split}
\end{align*}
After that, we can write for the dissipative part of the motion as before,
\begin{align*}
	\begin{split}
		r^t = & -1 + 2\exp(\left(\eta^t_+-\eta^c_+\right)), \text{ for $\lambda_+$, and}\\
		r^t = & 1 - 2\exp(\left(\eta^t_--\eta^c_-\right)), \text{ for $\lambda_-$}.\\
	\end{split}
\end{align*}
Thence,
\begin{equation}\label{rtime}
	r^t = \left(\exp(\eta^t_+-\eta^c_+)-\exp(\eta^t_--\eta^c_-)\right),
\end{equation}\\
and also,
\begin{equation*}
	\implies \lambda_{\pm}(t) = \exp(\eta^t_{\pm}-\eta^c_{\pm}).
\end{equation*}
Where, $\eta^t_{\pm} = \left(\ln(\lambda_{\pm}')+(\beta_H H_{\pm}-\beta_I)\right)e^{-\frac{t}{\tau}}$, $ \lambda' $ as in equation (\ref{initial}). The full evolution is given by,
\begin{equation}\label{qubitevol}
	\mathcal{U}_t\mqty[\dmat{\exp(\eta^t_+-\eta^c_+),\exp(\eta^t_--\eta^c_-)}]\mathcal{U}_t^{\dagger}.
\end{equation}
This solution above in equation (\ref{qubitevol}) works when we have Lagrange multipliers fixed using initial conditions \textit{i.e.,} FLM method. Otherwise, in general $\beta_i$'s depend on time-dependent $r$ and on constant $r_e=\hat{h}\cdot\vec{r}$. Consequently, the above equation (\ref{rtime}) needs to be solved numerically. For a detailed analysis, start from the differential equation given in the Appendix \ref{Appendix2}, equation (\ref{seaqa3}).

\begin{figure}[b]
	\centering
	\includegraphics[width=0.5\textwidth]{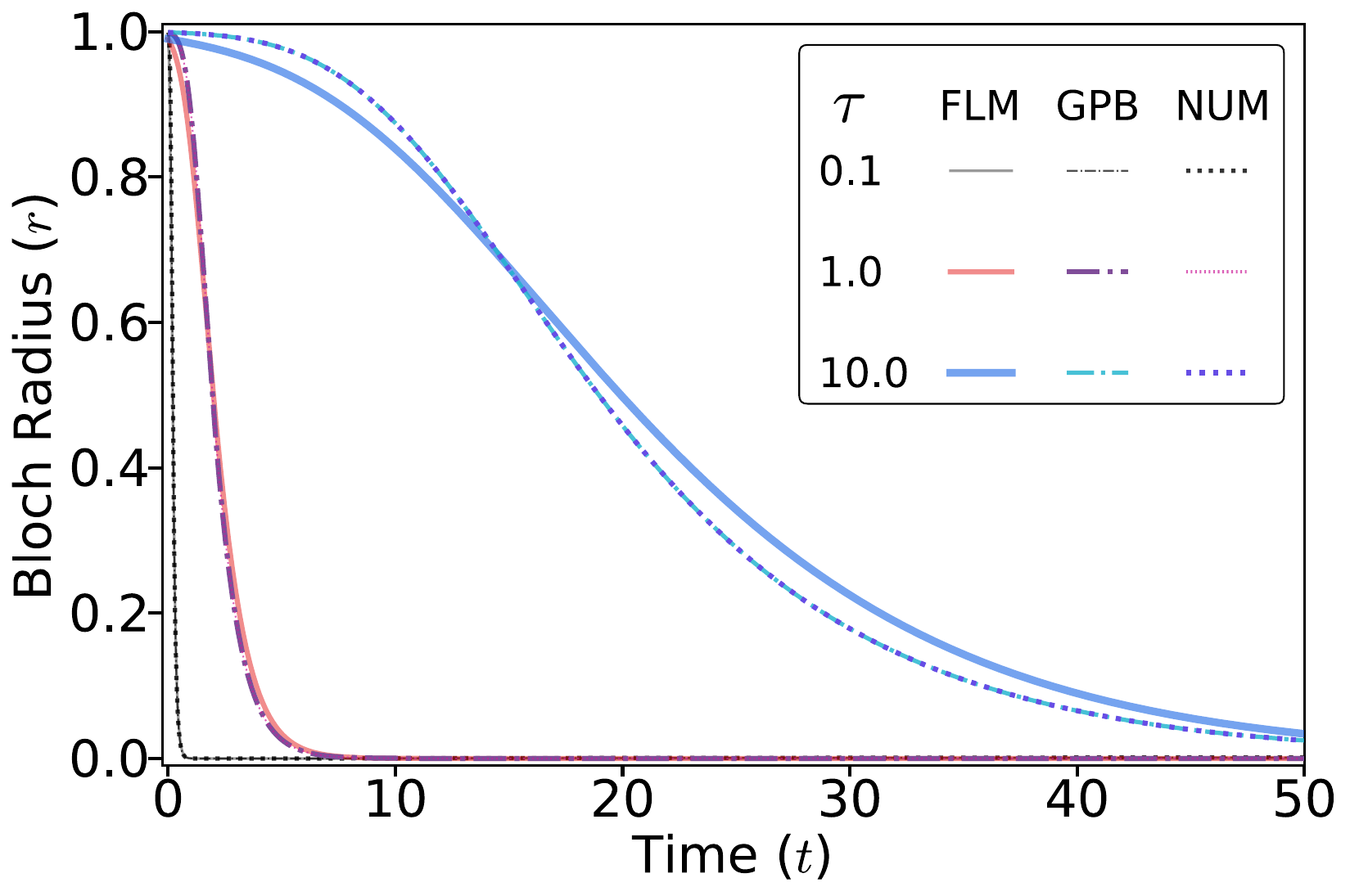}
	\caption{\label{2dradial}(Color online) Plot of the radial distance of the state operator in the Bloch sphere over time ($t=50$) for a qubit under various $\tau$ (first column of the legend), with $\omega=5$, and $\varepsilon=0.999$. As discussed in the text, the higher $\tau$ states fall slowly compared to lower valued ones. Solid lines represent computation done using FLM, dot-dash lines GPB are plotted using equation (\ref{berettaequation}) \cite{Beretta1985}, and the dotted NUM lines are plotted using direct numerical simulation of equation (\ref{req}).}
\end{figure}

\begin{figure}[h]
	\centering
	\subfigure[\label{2dspiral}]{\includegraphics[width=0.95\linewidth]{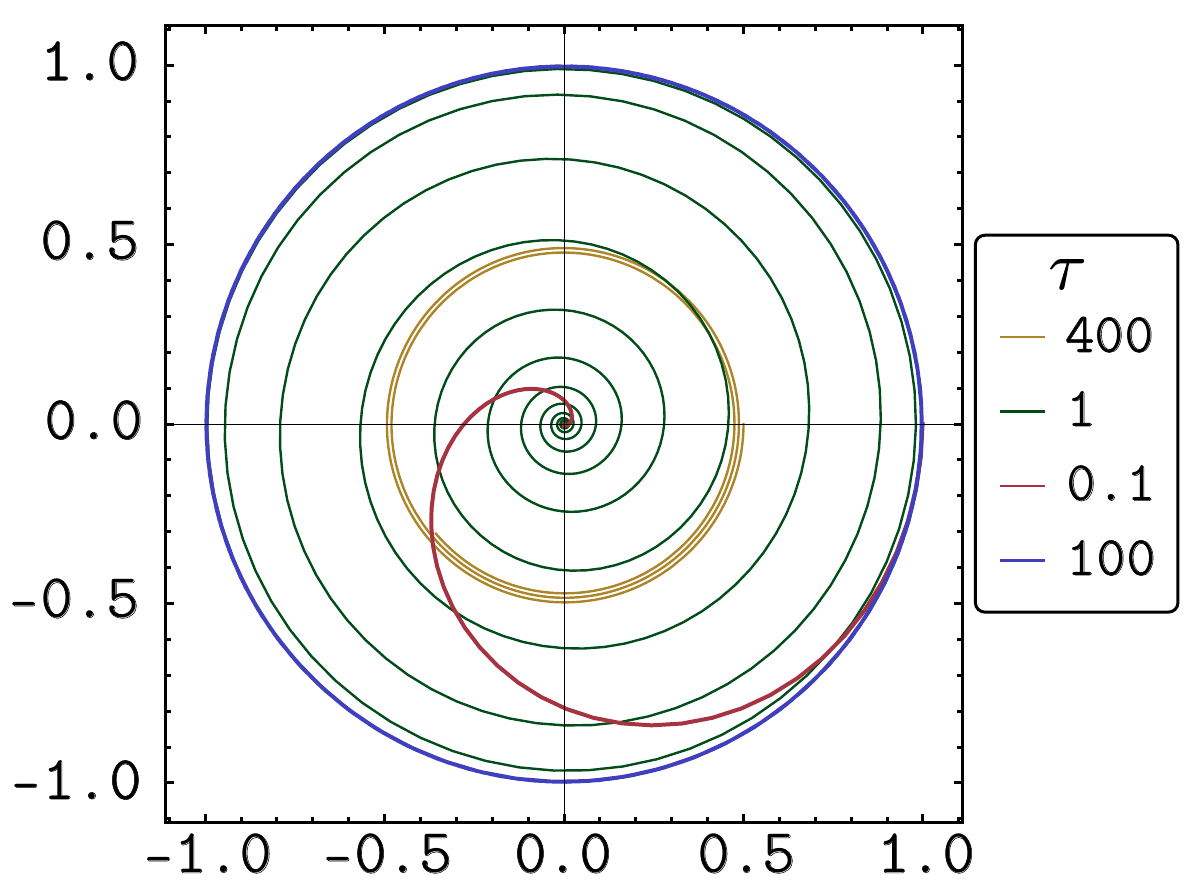}}
	\subfigure[\label{entropyspiral}]{\includegraphics[width=0.95\linewidth]{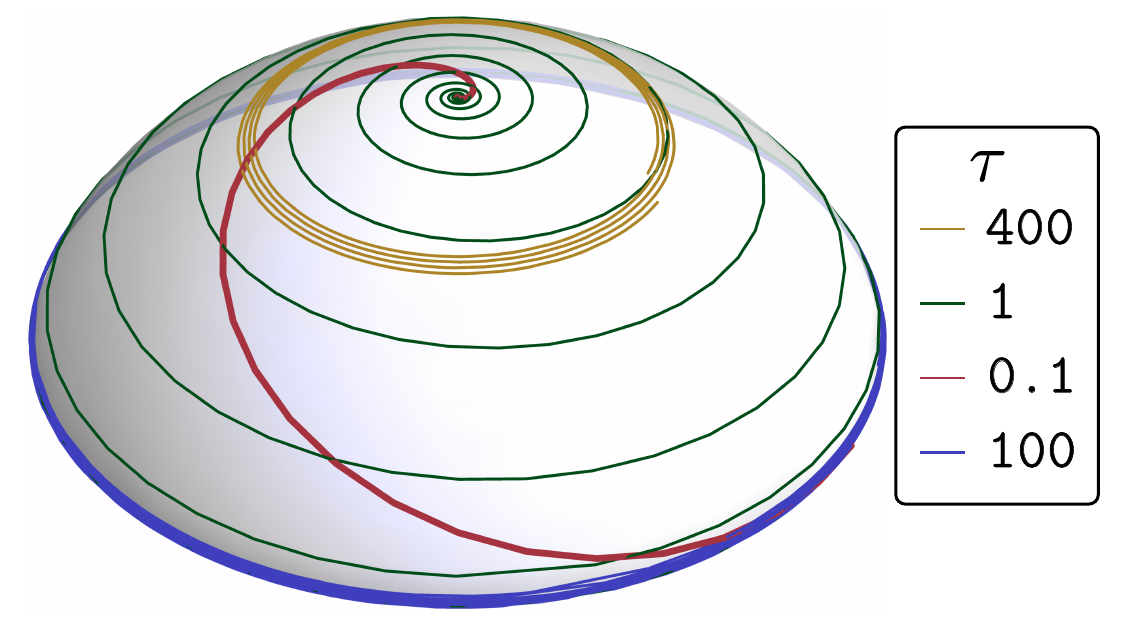}}
	\caption{\label{spiralmain}(Color online) (a) The spiraling trajectory of the state operator in the equatorial plane of the Bloch sphere, as viewed from its North pole after $t=30$. Different $\tau$ values show the difference between each trajectory. (b) Evolution of states after $t=30$, on the surface of revolution generated from the entropy functional expression. The legend demarks the $\tau$ values considered for this plot. The case with $\tau = 400$ correponds to the case with $\varepsilon = 0.5$, while the rest of the cases have $\varepsilon = 0.999$. Lower $\tau$ states rise faster, and the lowest has the steepest ascent.}
\end{figure}

Let us now understand the SEA approach through simple well-known, and well-studied physical conditions. We take $\hat{h} = \hat{z}$, and focus on the states lying on the equatorial plane of the Bloch sphere, $r_e=0$. $\mathcal{H}$ in this scenario becomes $\omega\sigma_z$, which is diagonal in the standard basis. Using the expression for $ \lambda' $ provided in equation (\ref{initial}), we write down the $\beta_i$'s as (equations (\ref{seaqbetah}), and (\ref{seaqbetai})),
\begin{align}\label{betass}
	\begin{split}
		\beta_H = &\quad 0,\\
		\beta_I = & \dfrac{k}{2}\left[\ln(\dfrac{1-\varepsilon^2}{4})+\varepsilon\ln(\dfrac{1+\varepsilon}{1-\varepsilon})\right].
	\end{split}
\end{align}
Let the initial $\rho$ be $a\dyad{0}{0} + b\dyad{1}{1}$. Then radius $ r' $ after quenching is $r' = \varepsilon r = \varepsilon\abs{a-b}$. We get, $\eta^t_{\pm} = \left(\ln(\frac{1\pm\varepsilon}{2})-\beta_I\right)e^{-\frac{t}{\tau}}$,
and $\eta^c_{\pm}= \beta_I$. And finally, the evolution equation becomes,
\begin{equation}\label{revolution}
	r^t = \left(e^{\theta_+^t}-e^{\theta_-^t}\right),
\end{equation}
where, $\theta_{\pm}^t = \eta^t_{\pm}-\eta_c{\pm}$. For $\omega = 5$, and $\varepsilon  = 0.999$ \footnote{just an arbitrary number close to one.}, we graphically show the evolution of $r^t$ in FIG. \ref{2dradial}.
As evident from the form of the general dynamical equation (\ref{eq1}), $\tau$ acts as modulating factor, where high $\tau$ results in a smaller dissipation contribution. Trivial mathematical interpretations aside, system relaxation time $\tau$ behavior is inversely related to the entropy production rate. After the system goes through a quenching process, it relaxes and could either thermalize or localize. This behavior is dependent mainly on the speed at which this happens. Higher $\tau$ implies slower relaxation, while as commented in literature, lower positive values of $\tau$ result in the steepest ascent of entropy, as we see in FIG. \ref{spiralmain}.
In the expression of $\eta_t$ equation (\ref{p_isol}), the exponent has ($\frac{-t}{\tau}$) dependence, which implies in $\frac{t}{\tau}<<1$ we will have a non-dissipative feature, and at larger times we will have the desired dissipation.

From the FIG. \ref{2dradial} we see higher $\tau$ valued states will have more delayed and gradual relaxation. To coroborate our results with the existing ones in the literature, we consult the solution provided in Ref. \cite[equation (19)]{Beretta1985}(GPB), reproduced below:
\begin{equation}\label{berettaequation}
	r^t =  \tanh(\half\exp(-\frac{t}{\tau})\ln(\frac{1+\varepsilon}{1-\varepsilon})).
\end{equation}
Finally, we have plotted the numerical solution (NUM) of equation (\ref{req}). We can see that the GPB solution matches the NUM results. FLM values, on the other hand, lie somewhat close, yet the initial and final agreements between FLM and GPB/NUM are intriguing, given that FLM is an approximation. Depending on the $\beta_i$ considered, FLM can be fine-tuned. If we fix $\beta_i$ at initial values, FLM better represents far from equilibrium behavior and equilibrium behavior FIG. \ref{2dradial}. However, FLM tuned using the final value of $\beta_i$ does a better job representing near equilibrium feature but doesn't fare equally well in far from equilibrium region. Moreover, as $\tau$ changes, disagreement between the FLM and GPB/NUM increases.

\begin{figure}[h]
	\centering
	\includegraphics[width=0.95\linewidth]{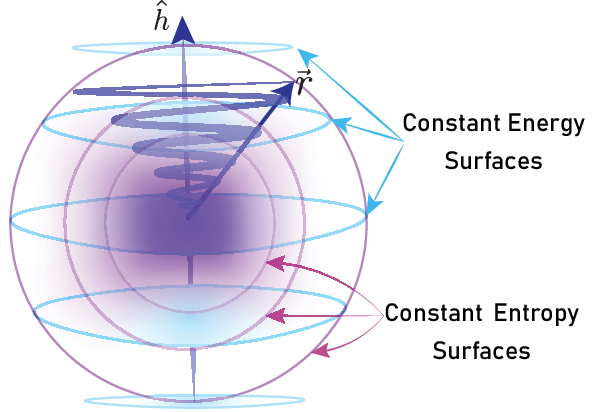}
	\caption{\label{spirality} (Color online) A schematic depiction of a qubit state's general energy dissipative evolution (purple spiral in the online colored version). This schematic is for high $\tau$ values; for low values of the same, the trajectory will follow a straighter path, as indicated in Ref. \cite{vallejo2021}, which will also be the steepest ascent. }
\end{figure}

We show the spiraling motion to the center of the Bloch sphere on the equatorial plane in FIG. \ref{2dspiral}. Here, we see that high $\tau$ states remain near the pure states for a longer time than low-valued ones, as they almost instantaneously mix to the maximum entropic state. These low values of $\tau$ trajectories represent the steepest entropy ascent solution. This steep \textit{ascent} can be better visualized when we consider the surface of revolution generated from the entropy functional in equation  (\ref{entropyonr}) and plot these spiral trajectories onto that surface as shown in FIG. \ref{entropyspiral}. We see that with time each trajectory tries to rise to the top of the surface where the point with maximum entropy is present. We can also see that high $\tau$ states maintain a limiting behavior at the foot of the surface of revolution, taking almost forever to reach the top (unitary type behavior). A schematic evolution of the general non-energy conserving motion for a qubit under SEA is presented in FIG. \ref{spirality}.
As energy keeps on decreasing, the value of $r_e$ also decreases. The state arrives at the global equilibrium when it reaches zero at the center. Next, we consider the evolution of a walker performing continuous-time quantum walk under spontaneous decoherence.
%....................................................................................................................
\subsection{\label{qudit}Case II - $N$-level system: Quantum Walker}
As done in the previous sub-section, we can very well use a Bloch vector representation for $d = N$ dimensional qudit state space as a $\mathbb{R}^{N^2-1}$ dimensional ball\cite{Park1971I, Park1971II, KIMURA2003339,bertlmann2008bloch,ozols2020generalized}, with a unit vector $\vec{r}$ associated with the space. We write $ \rho $ as\cite{ozols2020generalized},
\begin{equation}\label{blochgen}
	\rho = \frac{1}{N}\left(\mathrm{I}+\sqrt{\frac{N(N-1)}{2}}\vec{r}\cdot\vec{\Gamma}\right).
\end{equation}
Where $\vec{\Gamma}$ is a Pauli type generalization in $N$ dimension, comprising of generators of $SU(N)$ along with the identity operator $\mathrm{I}$\cite{KIMURA2003339}. Yet this is not a trivial generalization of the familiar Bloch sphere representation because in dimensions $N\ge3$, the expression (\ref{blochgen}) doesn't offer a bijective mapping like the qubit case. In the qubit scenario, each point on and inside the Bloch sphere has a one-to-one correspondence with a physical state, referring to a semi-positive density matrix with unit trace. We sometimes get non-physical results (holes) in the generalized Bloch sphere, which renders any trivial generalization ineffective \cite{bertlmann2008bloch}. The radius of this generalized Bloch sphere is given by $r = \sqrt{\frac{N}{N-1}(R-\frac{1}{N})}$ \cite{ozols2020generalized}, where $R=\tr(\rho^2)$.

\begin{figure}[h]
	\centering
	\includegraphics[width=0.5\textwidth]{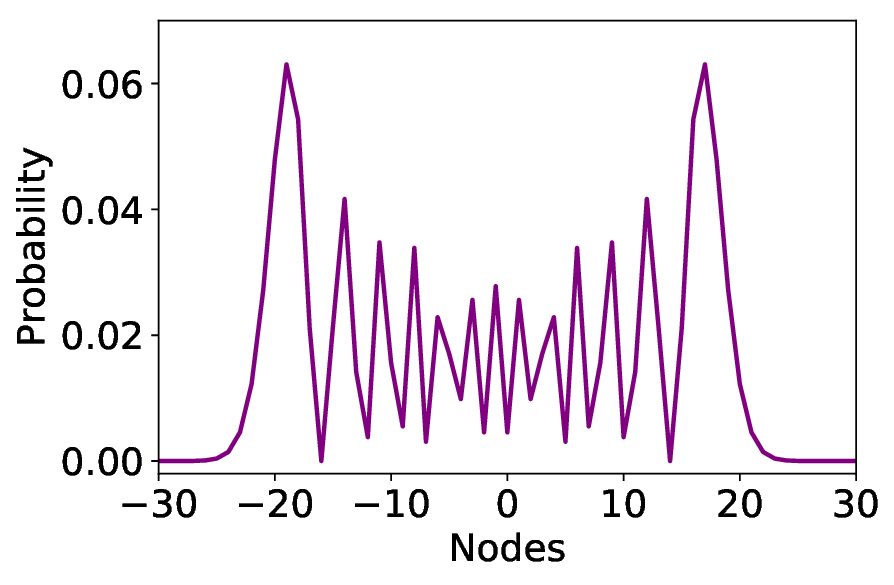}
	\caption{\label{unitaryctqw}(Color online) Probability distribution for a single quantum walker on a ring of 100 nodes. The walker was initiated at node 50, and the probability distribution is after $t=10$. In this diagram, node 50 is shifted to node 0.}
\end{figure}

The usual probability distribution computed from equation (\ref{qwrho}) for a walker performing CTQW on a ring of $N = 100$ nodes after some time $t=10$ is shown in FIG. \ref{unitaryctqw}, which is similar to CTQW walk distribution on a line for a short time and large $N$.\footnote{It is to be noted while we find probability distributions such as this at some time $t$, they are not the same for the discrete and continuous case. In the continuous case, we have transition amplitudes per unit time ($\mu$), and we consider probabilities at some instance post-initiation. These instances can be recorded as \textit{steps}. Wherein for the discrete case, a coin operation followed by a swap operation constitutes a \textit{step}.} For simplicity, we have considered $\mu=1$ \cite{Manouchehri2014}, which means an unbiased transition to any adjacent vertex in an undirected graph $\mathcal{G}$ with no loops \footnote{In the FIG. \ref{unitaryctqw}, we have only shown the nodes up to which the walker has spread after t=10; it does not show all nodes.}.
The density matrix of the walker at any time can be described using the standard basis as
\begin{equation}
	\rho^t = \sum_{i}p^t_i\dyad{i}{i},
\end{equation}
where $p^t_i$ is the probability of finding the walker on a node (vertex) $i$ after some time $t$ since the walk was initiated. As we have already described, the laplacian of $\mathcal{G}$ can be expressed as
\begin{align}\label{laplacianexpand}
	\begin{split}
		\mathbf{L} = & \mathbf{D} - \mathbf{A} \\
		= & \sum_{ij}\left(d_i\delta_{ij}\dyad{i}{j}-\mathbb{E}_{ij}\left(\dyad{i}{j}+\dyad{j}{i}\right)\right),
	\end{split}
\end{align}
where, $\mathbb{E}_{ij} = 1$ when there is an edge element $e_{ij}\in\mathbb{E}$, and zero otherwise. We notice from the Hamiltonian presented in the standard basis; the diagonal elements are simply the degree matrix entries, $H^d_{ii} = \mu d_{i}$, $d_i$ is the degree at the vertex $v_i$. So in the case of these walks, using the computation of $\beta_i$ carried out in Appendix \ref{Appendix4}, the equation (\ref{fullevolution}) becomes,
\begin{align}\label{ctqwevolution}
	\begin{split}
		&\rho^t = \\ &\exp(-i\mathcal{H}t)\left(\sum_{m}\exp(\eta^t_m-\eta^c_m)\mathbb{P}_m\right)\exp(i\mathcal{H}t),
	\end{split}
\end{align}
where, $\eta^t_m = \left(\ln(p_m^0)+\eta^c_m\right)e^{-t/\tau} = \tilde{v}_i\eta_t$, and $\eta^c_m = \mu\beta_H d_i-\beta_I$, and $\mathbb{P}_m$ is the projection operator as before. $p_m^0$ is found using equation (\ref{initial}).
\begin{figure}[h]
	\centering
	\includegraphics[width=0.5\textwidth]{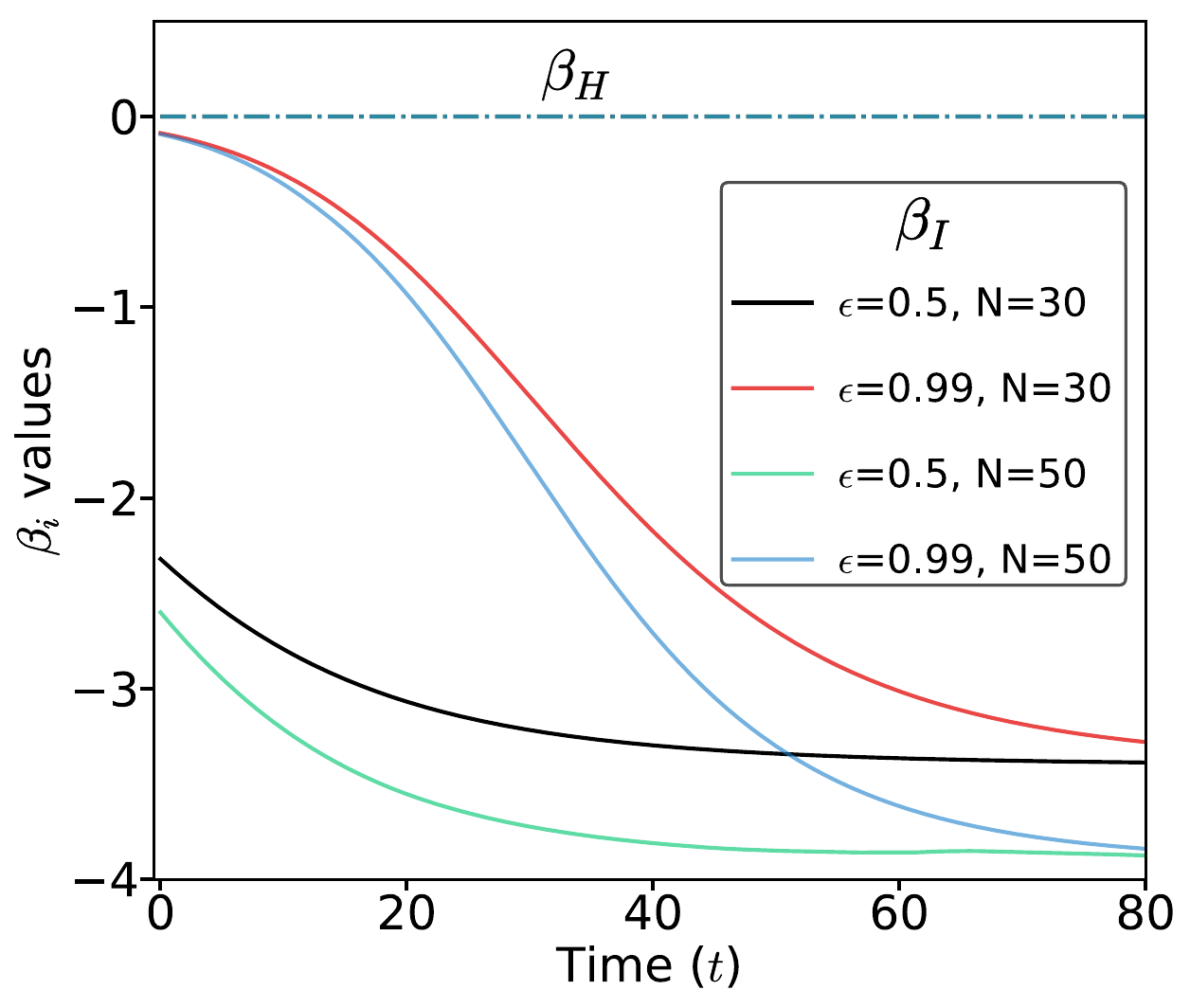}
	\caption{\label{betai}(Color online) Plot of $\beta_i$ vs time. In the legend, $ \beta_I $ is tagged, as it varies for different $ N $ and $ \varepsilon $ values. $\beta_H$ is the constant line $ y=0 $ .}
\end{figure}
We use the predefined parameters to model the walk to understand what is happening and whether the solution (\ref{ctqwevolution}) is meaningful. A cycle graph $\mathcal{C}_N$ being 2-regular, has the following Hamiltonian in the standard basis,
\begin{equation}\label{hamilctqw}
	\mathcal{H} = \sum_{i}\left(2\dyad{i}{i} - \mathbb{E}_{i,\tilde{i}}\left(\dyad{i}{\tilde{i}}+\dyad{\tilde{i}}{i}\right)\right),
\end{equation}
where $\tilde{i} = i \pmod{N} + 1$. Using this $\mathcal{H}$, and the equilibrium distribution $ \rho_u $, we get from equation (\ref{ctqwbetas}) the limiting $\beta_i$ values which we use for CTQW (here we are interested in the equilibrium behavior),
\begin{align}\label{cqwbetanumerical}
	\begin{split}
		\beta_H = &~ 0,\\
		\beta_I = & -\ln(N),
	\end{split}
\end{align}
where $k=1$. In FIG \ref{betai}, we plot the variation of $\beta_I$ for two different $N$ values of 50 and 30, respectively. We numerically solve the equation (\ref{req}) and use the $\rho$ thus produced at each iteration to compute $ \beta_i $'s as defined in Appendix \ref{Appendix2}. This plot suggests that $\beta_I$ assumes a final value dependent on $N$ and the mean energy. Consider the red and black lines, for instance. As time progresses, we see that they merge towards a fixed value which is given by equation (\ref{cqwbetanumerical}), suggesting that even though there is an initial dependence on $ \varepsilon $, as equilibrium approaches, all of the $ \beta_I $'s assume same value. Considering the initial $\beta_i$ in FLM will be prudent if one wishes to study the behavior far from equilibrium. Otherwise, fixing the multipliers at an equilibrium distribution to use for FLM will faithfully represent equilibrium behavior. The $\beta_H$ plot is shown by the $ y=0 $ line in the graph, which remains constant in this case. The variation in $ \beta_I $ lies within a single order of magnitude and does not reflect a strong difference in probability amplitudes, as seen in FIG. \ref{fig2}.
\begin{figure}[h]
	\centering
	\includegraphics[width=0.48\textwidth]{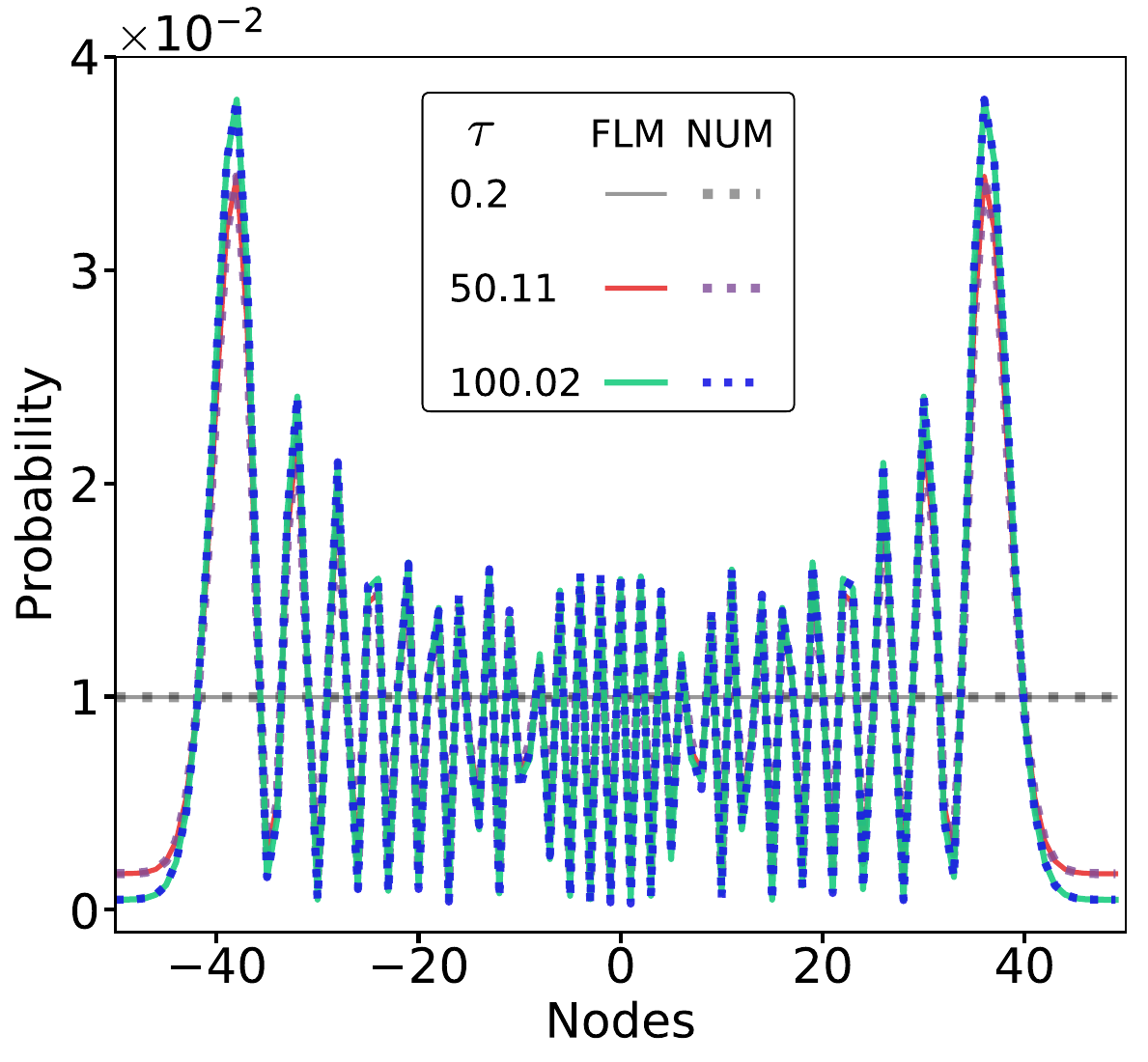}
	\caption{\label{fig2}(Color online) Probability distribution for a single quantum walker on a ring of 100 nodes under SEA conditions, initiated at node 50 (shifted to node 0 for symmetry), after $t=20$. The relaxation time $\tau$ is $0.2, 50.11$, and $100.02$ (first column in the legend), respectively. $\varepsilon$ is $0.99$. The plot contains both results from plotting the analytic solution found using the FLM method (solid lines in the plot) and those (dotted lines in the plot above) from the numerical solution to the equation (\ref{req}) using CTQW Hamiltonian and other relevant substitutions.}
\end{figure}
Using appropriate $\tau$ values, we get the probability distributions as plotted in FIG. \ref{fig2}. Here it is not easy to discern between results from the FLM (solid lines) and those from the numerical solution of the equation (\ref{req}). The typical values of probability amplitudes lie within the order of $10^{-2}$ as seen in the plot. From our numerical computation, we've estimated the difference with FLM results, which is of the order $10^{-4}$ for low $\tau$, and of the order $10^{-3}$ for high $\tau$s. We see, for the distributions considered after $t=20$, a similarity in the behavior of probability emerges as in FIGs \ref{2dradial}, \ref{2dspiral}. Higher $\tau$ or states closer to unitary states tend to relax slower. For low enough $\tau$, the rapid relaxation of the system is observed in FIG. \ref{fig2}, and all initial information is lost. On the other hand, high $\tau$ states having lesser entropy generation rates drive the system toward unitary-like behavior. This can also be understood in terms of localization and delocalization of the walker. The probability distribution for the case of $\tau = 0.2$ in FIG. \ref{fig2} shows strong delocalisation. While in the same figure, because of $\tau=50.11, 100.02$, and $t=20 < \tau$, we can say decoherence is yet to set in. That is, it displays linear behavior. As understood so far, low $\tau$ results in more non-linear behavior. But how low? Unfortunately, the answer to such a question remains elusive \cite{Beretta_SEA_87}.
\begin{figure}[t!]
	\centering
	\subfigure[\label{fig4a}]{\includegraphics[width=0.45\textwidth]{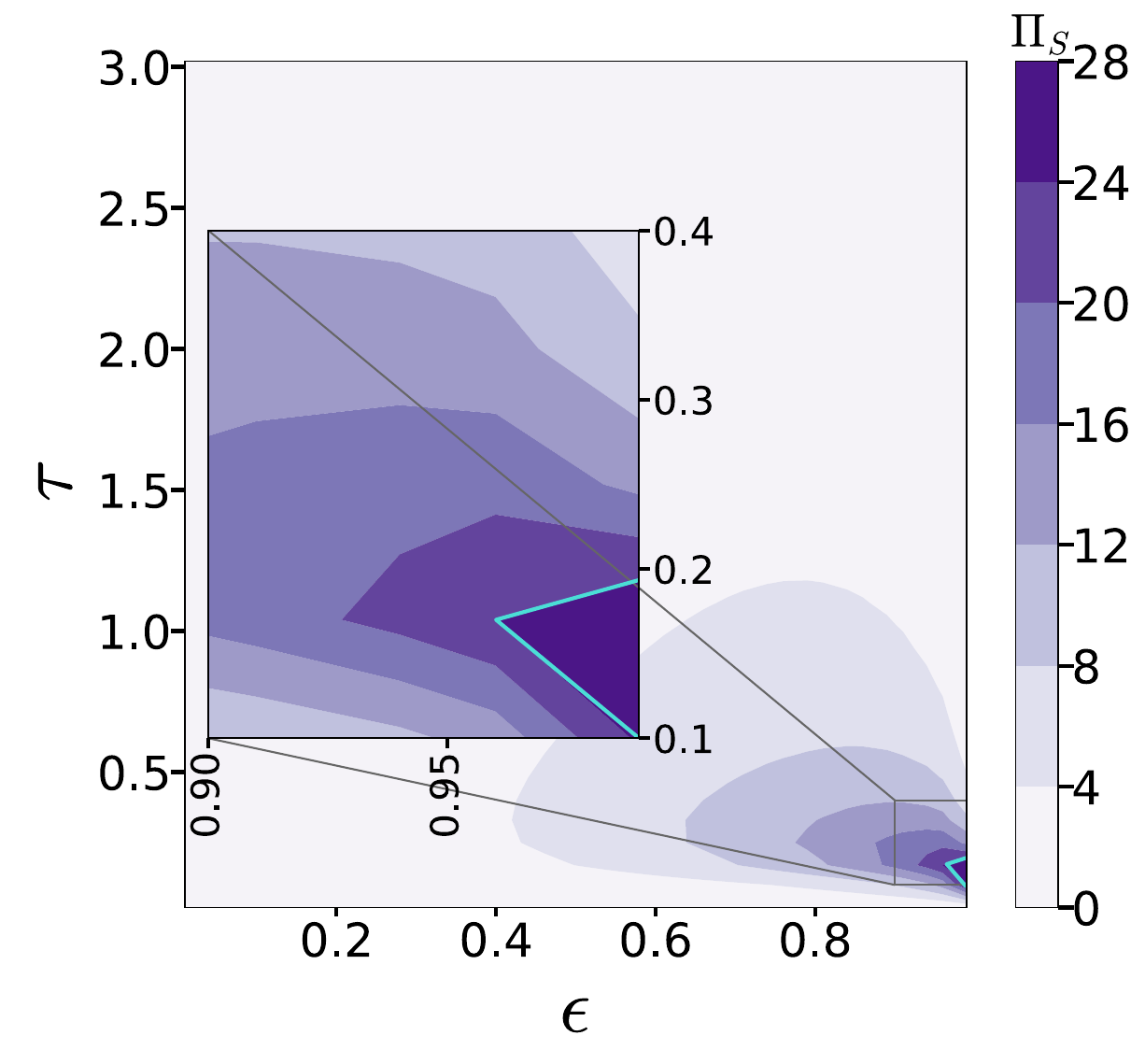}}
	\subfigure[\label{fig4b}]{\includegraphics[width=0.45\textwidth]{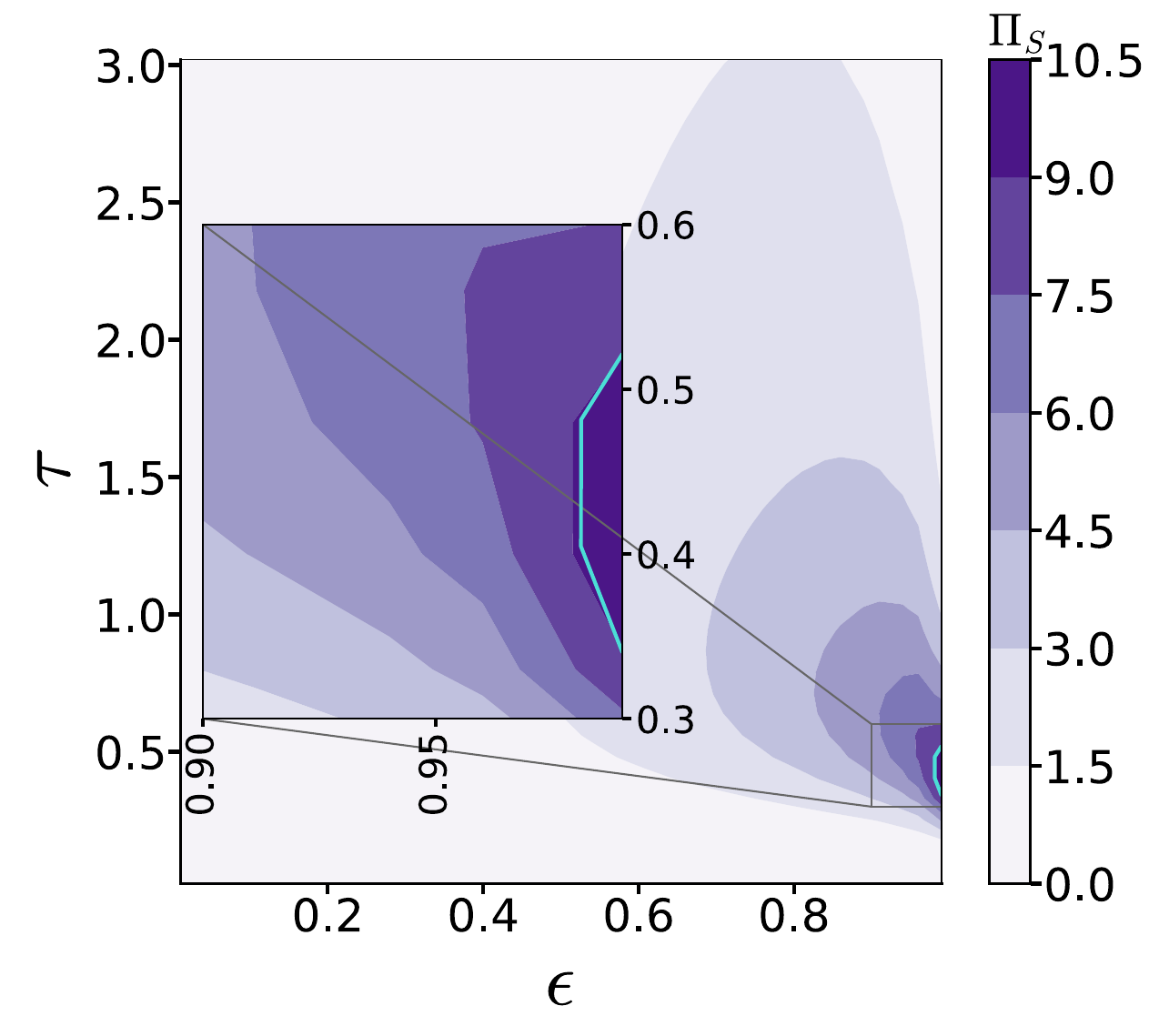}}
	\caption{\label{fig4}(Color online) $\Pi_S$ distribution for a CTQW of $N=50$ over different $\tau$ and $\varepsilon$ values after time $t=1$ in (a), and $t=3$ in (b). The color bars provide the range and contrast of $\Pi_S$ values. As discussed in the main text, the high $\Pi_S$ valued zones are concentrated around high $\varepsilon$ and low $\tau$ values. These deep purple areas bounded in cyan represent the maximum entropy generation area. In the insets of panels (a) and (b): zoomed-in view of the bounded region displaying max $\Pi_S$.}
\end{figure}
Yet, we can try to identify such low $\tau$ domains of high entropy generation by plotting the rate of entropy generation $\Pi_S$ (\ref{entropygeneration}) against $\tau$ and $\varepsilon$ values as given in FIG. \ref{fig4} for a CTQW with $N=50$. The leading contribution in determining $\Pi_S$ comes from $\tau$; one can confirm from FIG. \ref{fig4} that higher relaxation time represents almost zero entropy generation agreeing with our previous results. Lower $\tau$ states produce higher $\Pi_S$ values early on, as seen in FIG. \ref{fig4a}, which is a typical SEA behavior. Also, as time progresses, these highest $\Pi_S$ states (bounded by cyan line in the plots) start moving up along the right side of the diagram as in FIG. \ref{fig4b}. At a later time, only large $\tau$ valued states are yet to equilibrate, resulting in a change in entropy. Note that the quantity $\varepsilon$ by construction represents how pure the initial state is. So it is expected for a state with low $\varepsilon$ to reach equilibrium (that is, a noisy state becomes noisier) rather fast. Yet, the entropy production may not be maximum in those cases, as seen in FIG. \ref{fig4}. States relatively close to the pure states show maximum entropy production rate under SEA evolution (the deep purple shaded contours in the diagram). This behavior should be attributed to the fact that in the Bloch sphere representation, the low entropic states lying far away from equilibrium have to go through a greater change in entropy while equilibrating. Thus, their low information content accounts for a low entropy generation rate despite noisy channels becoming noisier.

\begin{figure}[t]
	\centering
	\subfigure[\label{Fig9_AvEn}]{\includegraphics[width=0.95\linewidth]{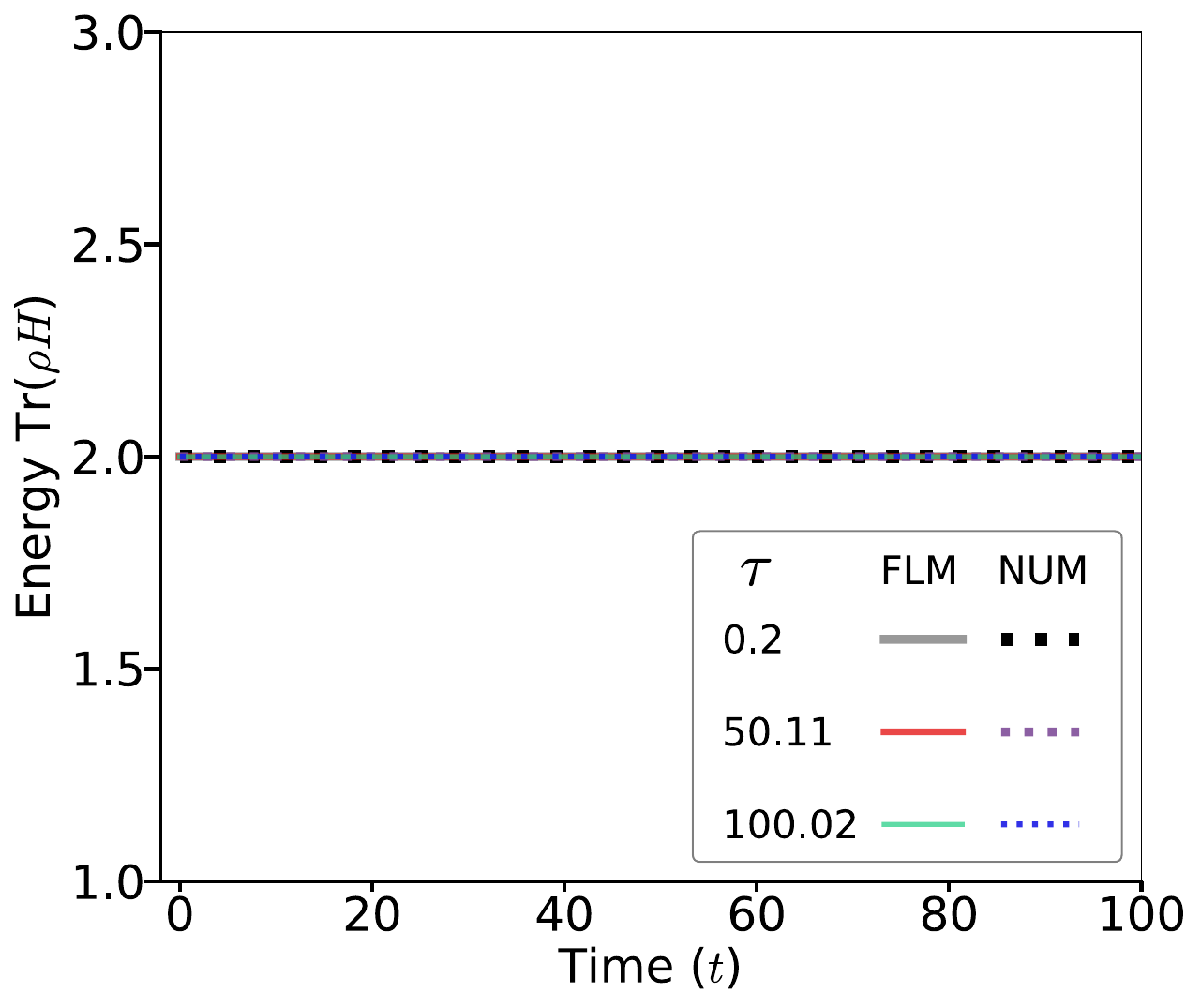}}
	\subfigure[\label{Fig9_S}]{\includegraphics[width=0.95\linewidth]{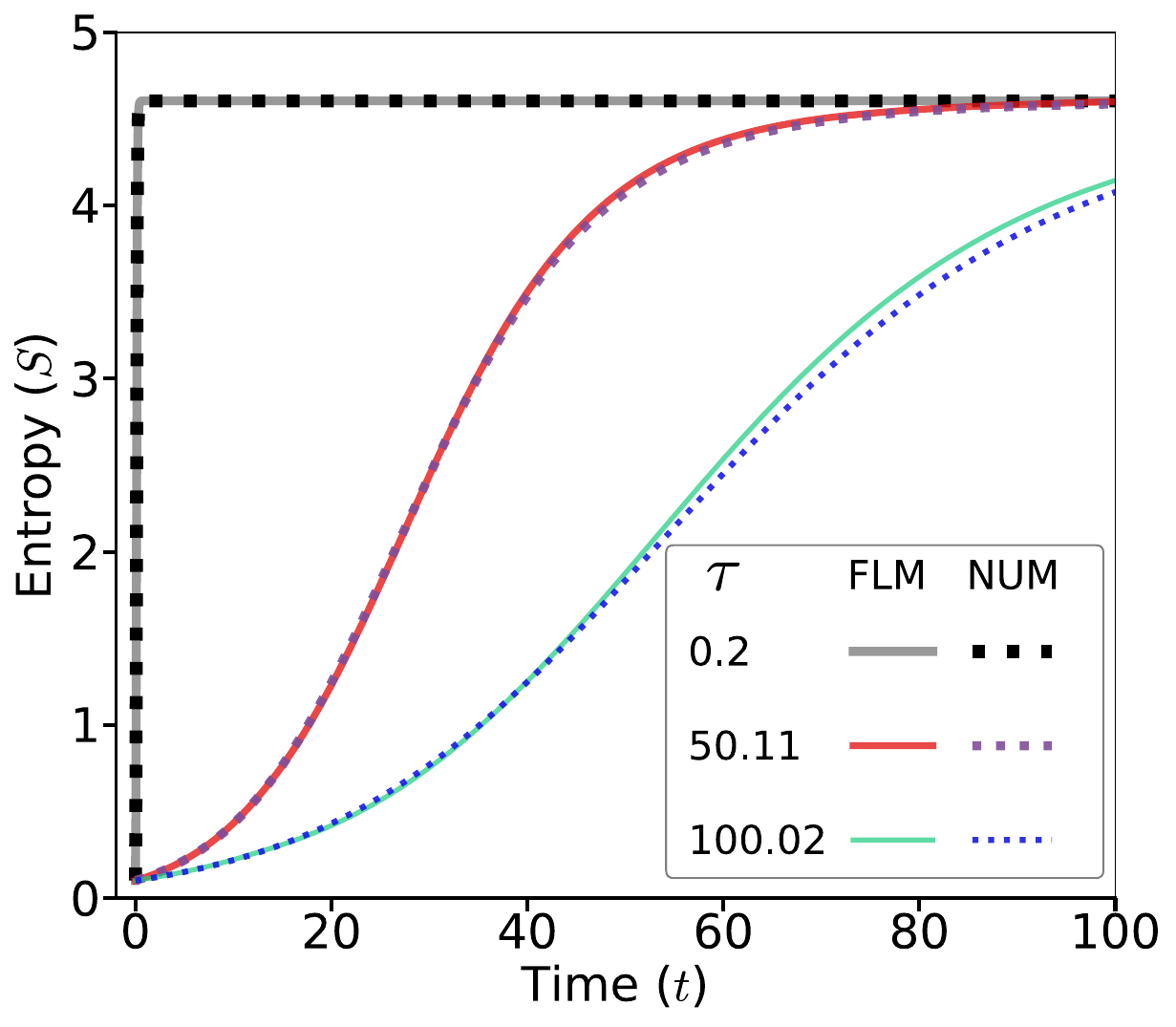}}
	\caption{\label{Fig9}(Color Online)(a) Plot of average energy vs time, and (b) entropy vs time for a CTQW on a cycle graph of 100 nodes for various $ \tau $ (first column of the legend) values and $ \varepsilon = 0.99 $. The walk was performed up to $ t=100 $. FLM (solid lines) denotes the analytically computed results, while NUM (dotted lines) denotes numerical results.}
\end{figure}

\begin{figure}[b]
	\centering
	\subfigure[\label{Fig10_PiS1}]{\includegraphics[width=0.95\linewidth]{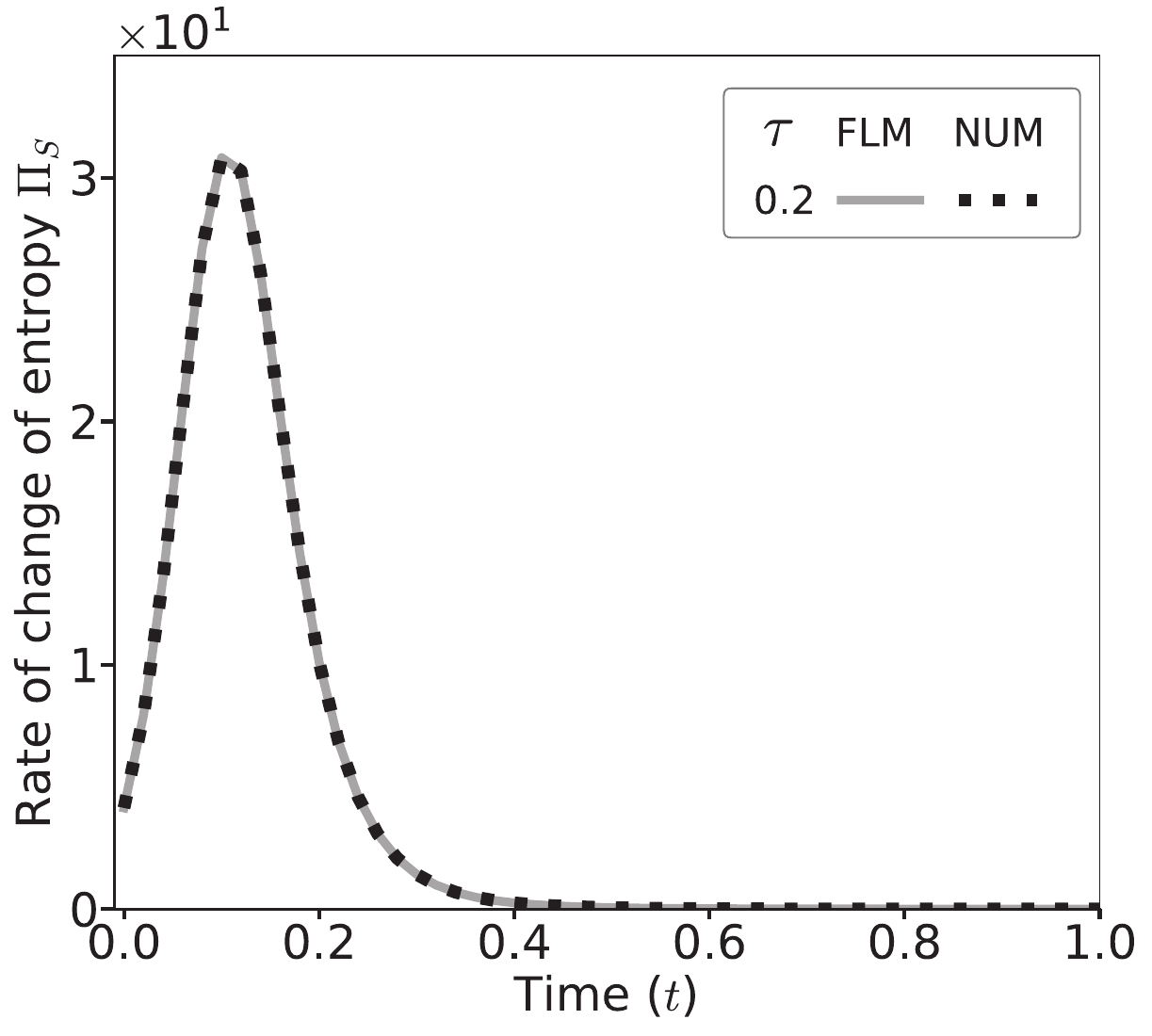}}
	\subfigure[\label{Fig10_PiS2}]{\includegraphics[width=0.95\linewidth]{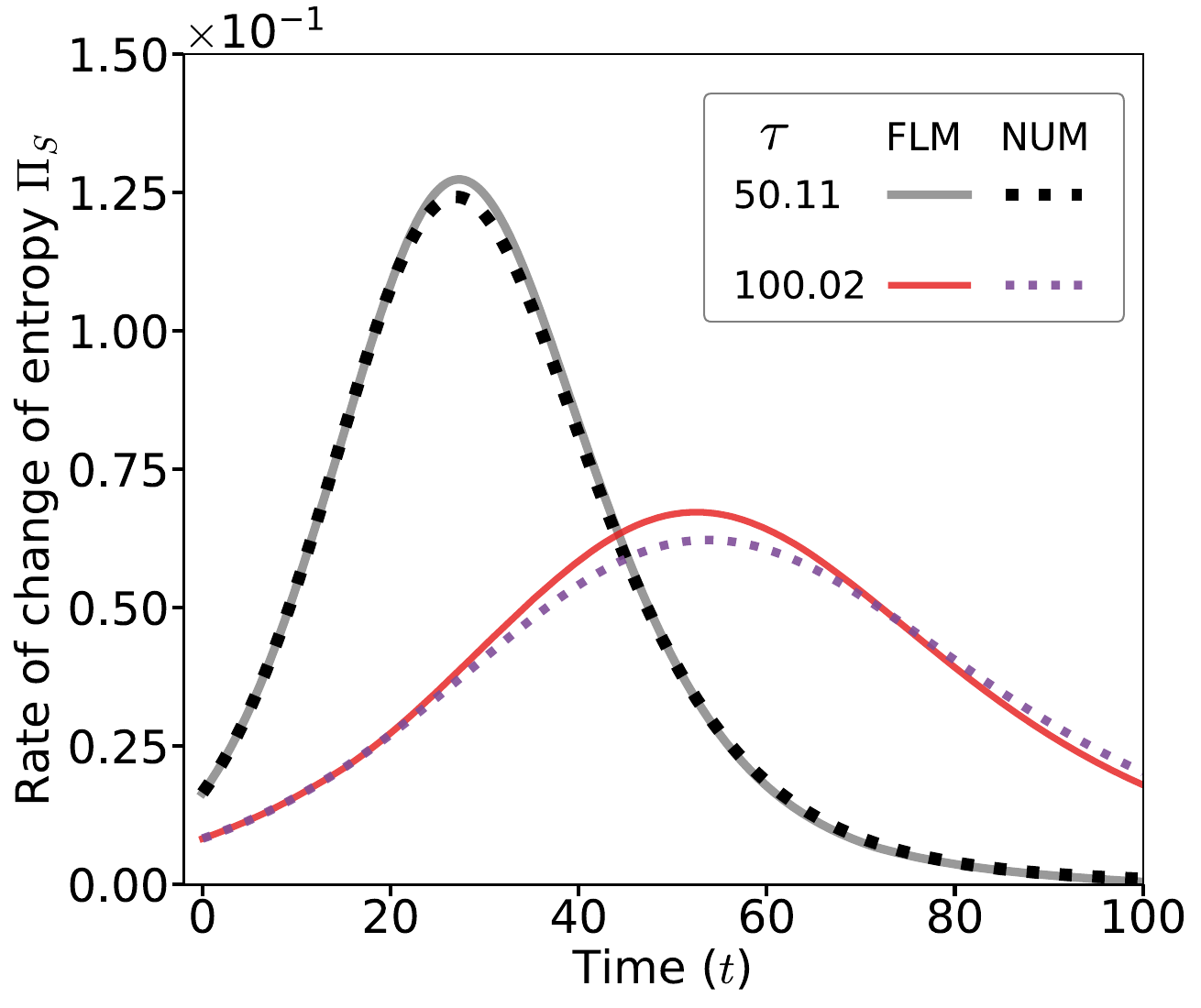}}
	\caption{\label{Fig10}(Color Online)Plot of rate of change of entropy vs time for a CTQW on a cycle graph of 100 nodes for (a) $ \tau = 0.2 $, and (b) $ \tau = (50.11,100.02) $ with $ \varepsilon = 0.99 $. FLM (solid lines) denotes the analytically computed results, while NUM (dotted lines) denotes numerical results.}
\end{figure}

We also run some consistency checks to validate our FLM results against numerical solutions of equation (\ref{seacompact}), such as computing average energy (FIG. \ref{Fig9_AvEn}), plotting entropy vs. time (FIG. \ref{Fig9_S}), and entropy generation rate vs. time as shown in FIG. \ref{Fig10}.
As before, we see a good agreement between the numerical and FLM results in FIG.s \ref{Fig9}, \ref{Fig10}. As the average energy remains constant throughout the walk (FIG. \ref{Fig9_AvEn}), we see entropy increasing monotonically and saturating at the maximum value (FIG. \ref{Fig9_S}). As we observe later, slight disagreements between full numerical (NUM in the plot) and FLM start appearing (blue dotted and green solid line in the plot). In this panel, we also see that almost instantaneously, the lowest $ \tau $ valued line (black and gray colored ones in the plot) reaches maximum entropy. To see the corresponding rate of change of $S$, we go to FIG. \ref{Fig10_PiS1}, where we see within $ t<0.1 $ the graph peaks around the value 32, which is twice the order of magnitude higher in other high $ \tau $ cases FIG. \ref{Fig10_PiS2}. This supports the SEA ansatz that the steepest entropic path is also the one with maximum entropy production rate. And this happens at low $ \tau $ values. Also, as noticed in FIG. \ref{2dradial}, as $\tau$ increases, we see differences between FLM and NUM results.

%%%%%%%%%%%%%%%%%%%%%%%%%%%%%%%%%%%%%%%%%%%%%%%%%%%%%%%%%%%%%%%%%%%%%%%%%%%%%%%%%%%%%%%%%%%%%%%%%%%%%%%%%%%%%%%%%%%%%%%%%%%%
\section{\label{conclusion}Discussions and conclusion}
Let us recapitulate what we have presented here. Our goal was to find spontaneously dissipative solutions to a continuous-time quantum walker (CTQW). In search of a suitable theory to characterize such dissipation, we arrive at the nonequilibrium thermodynamics propagated by Beretta and his co-workers under the name steepest entropy ascent (SEA) ansatz or local-SEA (LSEA) \cite{Beretta2020}. This theory entails a local description of the system and arrives at non-linear dynamics without invoking a heat bath or phenomenological tools such as presented by GENERIC dynamics. While these two approaches remain seemingly different in concept and application, in reality, it is not the case \cite{montefusco2015essential}. Hence, we choose SEA dynamics to govern dissipation in our CTQW system. The SEA principle can also be termed the fourth law of thermodynamics \cite{nonequilibriumthermodynamics}, which guarantees a unique global stable state with maximum entropy and all other states to be non-stable, meta-stable, or limit cycles. The SEA ansatz describes motion towards such a stable state under the local maximum entropy production principle by respecting the conservation laws. In that spirit, this work is a \textit{study of a single-particle quantum system as a consequence of the fourth law of thermodynamics}.

The non-linearity introduced in the general EoM of the SEA evolution renders any analytical attempt at a general solution highly difficult. Various special cases have been considered in literature till now. Herein, we provide a different approach by considering fixed Lagrange's multipliers (FLM) throughout the evolution under SEA within a good approximation of the numerical results. We have arrived at a solution that can be applied to varied cases. Our first example is a qubit. Despite being a simple system, it has been sparsely discussed in the SEA literature; we expound on that here. We have re-derived the solutions from our general equation (\ref{req}) using FLM on a Bloch sphere. From the magnitude of the Bloch vector vs. time plot, we see the effect of SEA dynamics in FIG. \ref{2dradial}. We observe a strong dependence of dissipation on the system's intrinsic relaxation time $\tau$. High $\tau$ meant if $t<<\tau$, the system is yet to relax, and unitary dynamics dominate the evolution, resulting in the quantum limit of the SEA dynamics. On the other extreme, for small but positive $\tau$, there is a rapid ascent towards the maximally mixed state, which is evident in the equatorial plot of FIG. \ref{2dspiral}. As the entropy generation rate is inversely related to $\tau$, this is also the domain of maximum entropy production. Here the system relaxes effectively instantaneously. Our approximate analytical results using FLM is an approximate result that agrees with the existing result due to \citet{Beretta1985} (GPB lines). The difference between GPB and FLM results in FIG. \ref{2dradial} originates from fixing multipliers. Although it is an approximate result, FLM works satisfactorily for two-level systems and $d$-level ones. FLM shows the agreement in FIG. \ref{2dradial}, giving us confidence in our approach.

We consider a similar analysis for SEA on a single CTQW, which is an $N-$level extension of the qubit case. In FIG. \ref{fig2} we compare our FLM results with the full numerical solution and find good agreement. We also plot the rate of entropy generation against $\varepsilon$ and $\tau$ and map the plot with contours. Low $\tau$ and high $\varepsilon$ cases showed a high rate of entropy generating contours, as seen in FIG. \ref{fig4}. We identify such regions of $\tau-\varepsilon$ where SEA dynamics dominate (see insets of FIG. \ref{fig4a}, and \ref{fig4b}). These are the states residing near the $\tr(\rho^2)=1$ showing maximum entropy generation in this figure. However, these marked domains in FIG. \ref{fig4} change with time, as different $\tau$ values activate $\Pi_S$ at different times, as seen in FIG. \ref{Fig10}. So, the goal of estimating a lower bound on $\tau$ has turned into a time-dependent problem. That is, for initial times, look at low $\tau$ for max entropy generation, and at later times look towards higher $\tau$ values. The agreement between FLM and numerical results continues to hold even when we compare average energy, entropy, and entropy growth over time in FIG.s \ref{Fig9}, and \ref{Fig10}. This shows the robustness of the FLM approach against various conditions.

\citet{kendonDecoherence2003} found decoherence to be helpful in quantum walks in their famous work. They introduced decoherence via weak measurements on each vertex. Instead, here, we do that via the SEA framework. We find that decoherence is useful in quantum walks. However, it can also be useful when the relaxation time of the system is short as it facilitates mixing. This paper shows that SEA with FLM can be used in many scenarios. This work raises further questions regarding the effect on QW's mixing time under such SEA dynamics. One may venture into the multi-walker case from here and seek to study various separability criteria that may be exploited.

%%%%%%%%%%%%%%%%%%%%%%%%%%%%%%%%%%%%%%%%%%%%%%%%%%%%%%%%%%%%%%%%%%%%%%%%%%%%%%%%%%%%%%%%%%%%%%%%%%%%%%%%%%%%%%%%%%%%%%%%%%%%

\section*{\label{acko}Acknowledgement}
I want to thank the Department of Science and Technology, Government of India, for the INSPIRE Fellowship. I also thank Prof. Sonjoy Majumder of the Department of Physics, Indian Institute of Technology Kharagpur, whose constant mentorship has been immensely valuable. I would also like to thank the anonymous referee of PRE, whose comments have helped shape and polish this manuscript and make it more robust. I acknowledge the National Supercomputing Mission (NSM) for providing computing resources of 'PARAM Shakti' at IIT Kharagpur, which is implemented by C-DAC and supported by the Ministry of Electronics and Information Technology (MeitY) and Department of Science and Technology (DST), Government of India.

%~~~~~~~~~~~~~~APPENDICES~~~~~~~~~~~~~~~~~~~~~~~~~~~~~~~~~~~~~~~~~

\appendix

\section{\label{Appendix1} Derivation of SEA EoM}

Here in this appendix, we sketch the basic derivation of the SEA equation of motion (EOM); by no means this is a complete formulation of what SEA pertains to (see the Refs in the main text for that purpose), but rather a guide to what is usually done.

\begin{figure}[h]
	\centering
	\includegraphics{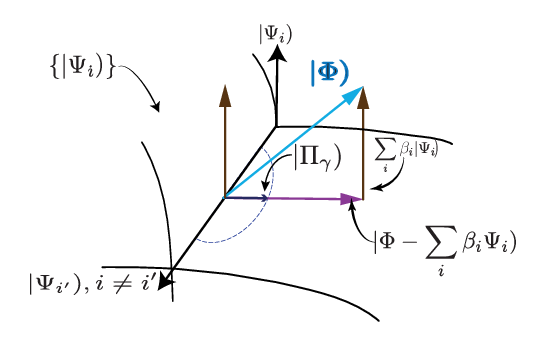}
	\caption{\label{geometry} (Color online) Geometric interpretation of the SEA principle. The manifold $\set{L}$ is spanned by the set $\left\{\kae{\Psi_i}\right\}$, which is shown in the figure as encoded on a surface, whose spanning vectors are shown along the edges. $\kae{\Phi}$ is the entropy gradient functional whose projection on the plane perpendicular to $\set{L}$ provides the direction of motion under SEA (purple arrow), while the magnitude of $\Pi_{\gamma}$ is constrained by a suitable value set to the equation (\ref{req11}), $\dv{l}{t} = \dot{\mathcal{\epsilon}}$, which defines a circle in the orthogonal plane, shown dotted in the diagram \cite{beretta2014steepest}.}
\end{figure}

The state operator $\gamma$'s rate of change follows the given equation \cite{Beretta5}:
\begin{equation}\label{appae1}
	\dot{\gamma} = \dot{\gamma}_H+\dot{\gamma}_D,
\end{equation}
Where $\dot{\gamma}_H$ represents evolution under purely Hamiltonian considerations and $\dot{\gamma}_D$ represents the same due to dissipation. While $\dot{\gamma}_H$ can be written using Schr\" {o} dinger's equation, the other component requires SEA formalism. Having defined the gradients in the main text (see equations (\ref{req3},\ref{req4})), we can associate the following geometrical notion to this motion. Let's consider FIG. \ref{geometry}: the conservation quantities and constraints of the motion span the linear manifold $\left\{\kae{\Psi_i}\right\}$, the entropy functional and $\gamma$ resides in the plane orthogonal to it \cite{beretta2014steepest}. The gradient of the entropy functional, the steepest ascent aspect, projects the two components, as shown in the figure. Our interest is in the component perpendicular to the manifold spanned by $\kae{\Psi_i}$s, denoted by the purple arrow in the diagram.
To fix the norm of the rate of change of $\gamma$, one includes the metric in the following fashion \cite{beretta2014steepest}.
\begin{equation}\label{req11}
	dl = \sqrt{\seaexp{\Pi_{\gamma}}{\hat{G}(\gamma)}{\Pi_{\gamma}}}dt.
\end{equation}
Thus, we have one more constraint.

Returning to the derivation, we can write the constraints using Lagrange multipliers in the following form,
\begin{equation}\label{req12}
	\Upsilon = \Pi_S - \sum_{i}\bar{\beta}_i\Pi_{C_i} - \frac{\tau}{2}\seaexp{\Pi_{\gamma}}{\hat{G}(\gamma)}{\Pi_{\gamma}},
\end{equation}
$\bar{\beta}_i$ and $\tau/2$ are Lagrange multipliers independent of $\Pi_{\gamma}$. The above equation can be reshaped using equations (\ref{req3}) and (\ref{req4}) and then on taking the functional derivative of $\Upsilon$ with respect to $\kae{\Pi_{\gamma}}$ we get,
\begin{equation}\label{req14}
	\frac{\delta \Upsilon}{\delta \Pi_{\gamma}} = \kae{\Phi} - \sum_{i}\bar{\beta}_i\kae{\Psi_i} - \tau\hat{G}(\gamma)\kae{\Pi_{\gamma}}.
\end{equation}
And the equation of motion for $\gamma$, $\dot{\gamma}_D$ to be precise, i.e., equation (\ref{reqm}) of the main text is found setting $\frac{\delta \Upsilon}{\delta \Pi_{\gamma}}=0$ as shown below.
\begin{equation}\label{appreqm}
	\kae{\Pi_{\gamma}} = \mathcal{L}\kae{\Phi - \sum_{i}\bar{\beta}_i\Psi_i}.
\end{equation}
$\mathcal{L}$ is given by $\frac{1}{\tau}\hat{G}(\gamma)^{-1}$, and it behaves as a superoperator with the properties $\mathcal{L}(A)B=\mathcal{L}(AB)$, and $(\mathcal{L}(A))^{\dagger}=A^{\dagger}\mathcal{L}$. For the purpose of our problem, we use $\mathcal{L}= \tfrac{1}{4k\tau}\mathrm{I}$ ($\mathrm{I}$ the Fisher metric, $k$ to counter the $k$ in entropy, and 4 for scaling). To derive equation (\ref{req}), we begin by using the following relation for dissipative motion
\begin{equation*}
	\dv{\rho_D}{t} = \Pi_{\gamma}\gamma^{\dagger}+\gamma\Pi_{\gamma^{\dagger}}=\dot{\gamma}_D\gamma^{\dagger}+\gamma\dot{\gamma}^{\dagger}_D.
\end{equation*}
We get using equation (\ref{reqm}) along with the equation above, together with the definitions- $\kae{\Phi}=\kae{\delta \mathbf{S}(\gamma)/\delta\gamma} = \kae{-2k(\ln{\gamma\gamma^{\dagger}}+1)\gamma}$ and $\kae{\Psi_i} = \kae{\delta \mathbf{C_i}(\gamma)/\delta\gamma} = \kae{2\mathbf{C_i}\gamma}$,  to get the following expression for the dissipative part of the dynamics,
\begin{align*}
	\begin{split}
		\Pi_{\gamma} = & -2\mathcal{L}\kae{k\ln(\gamma\gamma^{\dagger}+1)\gamma+\sum_{i}\bar{\beta}_i\mathbf{C_i}\gamma}, \\
		& \text{Using $ \dv{\rho_D}{t} = \Pi_{\gamma}\gamma^{\dagger}+\gamma\Pi_{\gamma^{\dagger}} $,} \\
		\dv{\rho_D}{t} = & -2\left[k\mathcal{L}(\ln(\gamma\gamma^{\dagger}))\gamma\gamma^{\dagger} +\sum_{i}\bar{\beta}_i\mathcal{L}(\mathbf{C_i})\gamma\gamma^{\dagger} + k\mathcal{L}\gamma\gamma^{\dagger} \right.\\
			&\left.+k\gamma\gamma^{\dagger}\mathcal{L} +k\gamma\gamma^{\dagger}\ln(\gamma\gamma^{\dagger})\mathcal{L} + \sum_{i}\bar{\beta}_i\gamma\gamma^{\dagger}\mathbf{C_i}\mathcal{L}\right].
	\end{split}
\end{align*}
Identifying $\gamma\gamma^{\dagger}=\rho$, we can rearrange the RHS of the last statement of the above equation to get the following one,
\begin{equation}\label{appaeD}
	\dv{\rho_D}{t} = -2\left[k\acomm{\mathcal{L}(\ln(\rho))}{\rho}+k\acomm{\mathcal{L}}{\rho} +\sum_{i}\bar{\beta}_i\acomm{\mathcal{L}(\mathbf{C_i})}{\rho}\right].
\end{equation}
Similarly, using Schr\"{o}dinger equation, we find
\begin{equation*}
	\dot{\gamma}_H = -\frac{i}{\hbar}\mathcal{H}\gamma,
\end{equation*}
Now, using $\dot{\rho} = \dot{\gamma}\gamma^{\dagger}+\gamma\dot{\gamma}^{\dagger}$, $\hbar=1$, and equations (\ref{appae1},\ref{appaeD}), we get the equation (\ref{eq1},\ref{req}) of the main text.

A useful term in our analysis is $\Pi_S$, the entropy generation rate functional from equation (\ref{req3}). It is given by,
\begin{align}
	\begin{split}
		\Pi_S =& -k\dv{\tr(\rho\ln(\rho))}{t},\\
		=& -k\tr(\left(\ln(\rho)+1\right)\dv{\rho}{t})\\
		=& 2k^2\tr(\left(\ln(\rho)+1\right)\mathcal{L}\acomm{\ln(\rho)}{\rho})\\
		&+ 2k^2\sum_{i}(-1)^i\beta_i\tr(\left(\ln(\rho)+1\right)\mathcal{L}\acomm{\mathbf{C_i}}{\rho}).
	\end{split}\tag{A7'}
\end{align}
The $\beta_i$ is scaled in the above equation (see equations (\ref{appBbetaM1}-\ref{appBbetaM3}) below). Using $\mathcal{L}= \frac{\mathrm{I}}{4k\tau}$, we get,
\begin{align}\label{entropygeneration}
	\begin{split}
		\Pi_S =& \dfrac{k}{2\tau}\Big(\tr((\ln(\rho)+1)\acomm{\ln(\rho)}{\rho})\\
		&+ \sum_i(-1)^i\beta_i\tr((\ln(\rho)+1)\acomm{\mathbf{C_i}}{\rho}) \Big).
	\end{split}
\end{align}
%---------------------------------------------------------------------------------------------------------------
\section{\label{Appendix2}Determining Lagrange multipliers}
Using the constraints of equations (\ref{req3}-\ref{reqm}) we get,
\begin{equation}\label{seaqa1}
	\sum_{i}\bae{\Psi_j}\mathcal{L}\kae{\Psi_i}\bar{\beta}_i = k\bae{\Psi_j}\mathcal{L}\kae{\Phi}.
\end{equation}
This equation can be solved using Cramer's rule for solving linear equation with multiple variables, provided the solution exists ( $\bar{\Omega}=\det(\Delta)\ne 0$), which is equivalent to the following expression (considering three constraints, and corresponding three $ \bar{\beta} $s):

\begin{align}\label{seaqa2}
	\begin{split}
		\bar{\Omega} = & \vmqty{\seaexp{\Psi_1}{\mathcal{L}}{\Psi_1} &  \seaexp{\Psi_1}{\mathcal{L}}{\Psi_2} & \seaexp{\Psi_1}{\mathcal{L}}{\Psi_3}\\
			\seaexp{\Psi_2}{\mathcal{L}}{\Psi_1} & \seaexp{\Psi_2}{\mathcal{L}}{\Psi_2} &\seaexp{\Psi_2}{\mathcal{L}}{\Psi_3}\\
			\seaexp{\Psi_3}{\mathcal{L}}{\Psi_1} & \seaexp{\Psi_3}{\mathcal{L}}{\Psi_2} & \seaexp{\Psi_3}{\mathcal{L}}{\Psi_3}}, \\
		\bar{\beta}_1 = & \dfrac{1}{\bar{\Omega}} \vmqty{\seaexp{\Psi_1}{\mathcal{L}}{\Phi} &  \seaexp{\Psi_1}{\mathcal{L}}{\Psi_2} & \seaexp{\Psi_1}{\mathcal{L}}{\Psi_3}\\
			\seaexp{\Psi_2}{\mathcal{L}}{\Phi} & \seaexp{\Psi_2}{\mathcal{L}}{\Psi_2} &\seaexp{\Psi_2}{\mathcal{L}}{\Psi_3}\\
			\seaexp{\Psi_3}{\mathcal{L}}{\Phi} & \seaexp{\Psi_3}{\mathcal{L}}{\Psi_2} & \seaexp{\Psi_3}{\mathcal{L}}{\Psi_3}}, \\
		\bar{\beta}_2 = & \dfrac{1}{\bar{\Omega}}\vmqty{\seaexp{\Psi_1}{\mathcal{L}}{\Psi_1} &  \seaexp{\Psi_1}{\mathcal{L}}{\Phi} & \seaexp{\Psi_1}{\mathcal{L}}{\Psi_3}\\
			\seaexp{\Psi_2}{\mathcal{L}}{\Psi_1} & \seaexp{\Psi_2}{\mathcal{L}}{\Phi} &\seaexp{\Psi_2}{\mathcal{L}}{\Psi_3}\\
			\seaexp{\Psi_3}{\mathcal{L}}{\Psi_1} & \seaexp{\Psi_3}{\mathcal{L}}{\Phi} & \seaexp{\Psi_3}{\mathcal{L}}{\Psi_3}}, \\
		\bar{\beta}_3 = & \dfrac{1}{\bar{\Omega}}\vmqty{\seaexp{\Psi_1}{\mathcal{L}}{\Psi_1} &  \seaexp{\Psi_1}{\mathcal{L}}{\Psi_2} & \seaexp{\Psi_1}{\mathcal{L}}{\Phi}\\
			\seaexp{\Psi_2}{\mathcal{L}}{\Psi_1} & \seaexp{\Psi_2}{\mathcal{L}}{\Psi_2} &\seaexp{\Psi_2}{\mathcal{L}}{\Phi}\\
			\seaexp{\Psi_3}{\mathcal{L}}{\Psi_1} & \seaexp{\Psi_3}{\mathcal{L}}{\Psi_2} & \seaexp{\Psi_3}{\mathcal{L}}{\Phi}}.
	\end{split}
\end{align}
On column rearrangement, we get:
\begin{align*}
	\begin{split}
		\bar{\beta}_2 &=  - \dfrac{1}{\bar{\Omega}}\vmqty{ \seaexp{\Psi_1}{\mathcal{L}}{\Phi} &\seaexp{\Psi_1}{\mathcal{L}}{\Psi_1} &  \seaexp{\Psi_1}{\mathcal{L}}{\Psi_3}\\
			\seaexp{\Psi_2}{\mathcal{L}}{\Phi} &\seaexp{\Psi_2}{\mathcal{L}}{\Psi_1} & \seaexp{\Psi_2}{\mathcal{L}}{\Psi_3}\\
			\seaexp{\Psi_3}{\mathcal{L}}{\Phi} &\seaexp{\Psi_3}{\mathcal{L}}{\Psi_1} & \seaexp{\Psi_3}{\mathcal{L}}{\Psi_3}}, \\
		\bar{\beta}_3 &=  \dfrac{1}{\bar{\Omega}}\vmqty{\seaexp{\Psi_1}{\mathcal{L}}{\Phi}& \seaexp{\Psi_1}{\mathcal{L}}{\Psi_1} &  \seaexp{\Psi_1}{\mathcal{L}}{\Psi_2} \\
			\seaexp{\Psi_2}{\mathcal{L}}{\Phi}&\seaexp{\Psi_2}{\mathcal{L}}{\Psi_1} & \seaexp{\Psi_2}{\mathcal{L}}{\Psi_2}\\
			\seaexp{\Psi_3}{\mathcal{L}}{\Phi}&\seaexp{\Psi_3}{\mathcal{L}}{\Psi_1} & \seaexp{\Psi_3}{\mathcal{L}}{\Psi_2} }.
	\end{split}
\end{align*}
Before explicitly finding out the $\bar{\beta}_i$s, let us consider equation (\ref{appreqm}), and substitute $\bar{\beta}_i$s to get the following equation \cite{beretta2014steepest},

\resizebox{.8\linewidth}{!}{
	\begin{minipage}{\linewidth}
		\begin{align}
			\begin{split}
				\Pi_{\gamma} &= \mathcal{L}(\Phi)\dfrac{\bar{\Omega}}{\bar{\Omega}}- \dfrac{\mathcal{L}}{\bar{\Omega}} \vmqty{\seaexp{\Psi_1}{\mathcal{L}}{\Phi} &  \seaexp{\Psi_1}{\mathcal{L}}{\Psi_2} & \seaexp{\Psi_1}{\mathcal{L}}{\Psi_3}\\
					\seaexp{\Psi_2}{\mathcal{L}}{\Phi} & \seaexp{\Psi_2}{\mathcal{L}}{\Psi_2} &\seaexp{\Psi_2}{\mathcal{L}}{\Psi_3}\\
					\seaexp{\Psi_3}{\mathcal{L}}{\Phi} & \seaexp{\Psi_3}{\mathcal{L}}{\Psi_2} & \seaexp{\Psi_3}{\mathcal{L}}{\Psi_3}}(\Psi_1) \\
				&+ \dfrac{\mathcal{L}}{\bar{\Omega}}\vmqty{ \seaexp{\Psi_1}{\mathcal{L}}{\Phi} &\seaexp{\Psi_1}{\mathcal{L}}{\Psi_1} &  \seaexp{\Psi_1}{\mathcal{L}}{\Psi_3}\\
					\seaexp{\Psi_2}{\mathcal{L}}{\Phi} &\seaexp{\Psi_2}{\mathcal{L}}{\Psi_1} & \seaexp{\Psi_2}{\mathcal{L}}{\Psi_3}\\
					\seaexp{\Psi_3}{\mathcal{L}}{\Phi} &\seaexp{\Psi_3}{\mathcal{L}}{\Psi_1} & \seaexp{\Psi_3}{\mathcal{L}}{\Psi_3}}(\Psi_2)\\
				&- \dfrac{\mathcal{L}}{\bar{\Omega}}\vmqty{\seaexp{\Psi_1}{\mathcal{L}}{\Phi}& \seaexp{\Psi_1}{\mathcal{L}}{\Psi_1} &  \seaexp{\Psi_1}{\mathcal{L}}{\Psi_2} \\
					\seaexp{\Psi_2}{\mathcal{L}}{\Phi}&\seaexp{\Psi_2}{\mathcal{L}}{\Psi_1} & \seaexp{\Psi_2}{\mathcal{L}}{\Psi_2}\\
					\seaexp{\Psi_3}{\mathcal{L}}{\Phi}&\seaexp{\Psi_3}{\mathcal{L}}{\Psi_1} & \seaexp{\Psi_3}{\mathcal{L}}{\Psi_2} }(\Psi_3),
			\end{split}\label{bankai} \\
			\begin{split}
				\Pi_{\gamma}&= \dfrac{\vmqty{\mathcal{L}(\Phi) & \mathcal{L}(\Psi_1) & \mathcal{L}(\Psi_2) & \mathcal{L}(\Psi_3)\\ \seaexp{\Psi_1}{\mathcal{L}}{\Phi}& 		\seaexp{\Psi_1}{\mathcal{L}}{\Psi_1} &  \seaexp{\Psi_1}{\mathcal{L}}{\Psi_2} & \seaexp{\Psi_1}{\mathcal{L}}{\Psi_3}\\ \seaexp{\Psi_2}{\mathcal{L}}{\Phi}&
						\seaexp{\Psi_2}{\mathcal{L}}{\Psi_1} & \seaexp{\Psi_2}{\mathcal{L}}{\Psi_2} &\seaexp{\Psi_2}{\mathcal{L}}{\Psi_3}\\ \seaexp{\Psi_3}{\mathcal{L}}{\Phi}&
						\seaexp{\Psi_3}{\mathcal{L}}{\Psi_1} & \seaexp{\Psi_3}{\mathcal{L}}{\Psi_2} & \seaexp{\Psi_3}{\mathcal{L}}{\Psi_3}}}{\vmqty{\seaexp{\Psi_1}{\mathcal{L}}{\Psi_1} &  \seaexp{\Psi_1}{\mathcal{L}}{\Psi_2} & \seaexp{\Psi_1}{\mathcal{L}}{\Psi_3}\\
						\seaexp{\Psi_2}{\mathcal{L}}{\Psi_1} & \seaexp{\Psi_2}{\mathcal{L}}{\Psi_2} &\seaexp{\Psi_2}{\mathcal{L}}{\Psi_3}\\
						\seaexp{\Psi_3}{\mathcal{L}}{\Psi_1} & \seaexp{\Psi_3}{\mathcal{L}}{\Psi_2} & \seaexp{\Psi_3}{\mathcal{L}}{\Psi_3}}}.
			\end{split}
		\end{align}
	\end{minipage}}

To find a more comprehensive expression, we make use of the following relations,
\begin{align}\label{traceRel}
	\begin{split}
		\seaexp{\Psi_i}{\mathcal{L}}{\phi} =& -\dfrac{1}{\tau}\left(\tr(\rho\half\acomm{\mathbf{C_i}}{\ln(\rho)})+\tr(\half\acomm{\mathbf{C_i}}{\rho})\right)\\
		\seaexp{\Psi_i}{\mathcal{L}}{\Psi_j} =& \dfrac{1}{k\tau}\tr(\rho\half\acomm{\mathbf{C_i}}{\mathbf{C_j}}).
	\end{split}
\end{align}
Using these relations, we find the expressions for $\bar{\beta}_i$ from equation (\ref{seaqa2}) as follows,
% \begingroup
% \allowdisplaybreaks
\resizebox{0.9\linewidth}{!}{
	\begin{minipage}{\linewidth}
		\begin{align}
			\begin{split}
				\bar{\Omega} =&  \dfrac{1}{(k\tau)^3}\vmqty{\tr(\tfrac{\rho}{2}\acomm{\mathbf{C_1}}{\mathbf{C_1}}) &  \tr(\tfrac{\rho}{2}\acomm{\mathbf{C_1}}{\mathbf{C_2}}) & \tr(\tfrac{\rho}{2}\acomm{\mathbf{C_1}}{\mathbf{C_3}})\\
					\tr(\tfrac{\rho}{2}\acomm{\mathbf{C_2}}{\mathbf{C_1}}) & \tr(\tfrac{\rho}{2}\acomm{\mathbf{C_2}}{\mathbf{C_2}}) &\tr(\tfrac{\rho}{2}\acomm{\mathbf{C_2}}{\mathbf{C_3}})\\
					\tr(\tfrac{\rho}{2}\acomm{\mathbf{C_3}}{\mathbf{C_1}}) & \tr(\tfrac{\rho}{2}\acomm{\mathbf{C_3}}{\mathbf{C_2}}) & \tr(\tfrac{\rho}{2}\acomm{\mathbf{C_3}}{\mathbf{C_3}})},
			\end{split}\label{appBomega} \\
			\begin{split}
				\bar{\beta}_1 = & -\dfrac{1}{k^2\tau^3\bar{\Omega}}\\
				&\times\vmqty{\tr(\tfrac{\rho}{2}\acomm{\mathbf{C_1}}{\ln(\rho)+1}) &  \tr(\tfrac{\rho}{2}\acomm{\mathbf{C_1}}{\mathbf{C_2}}) & \tr(\tfrac{\rho}{2}\acomm{\mathbf{C_1}}{\mathbf{C_3}})\\
					\tr(\tfrac{\rho}{2}\acomm{\mathbf{C_2}}{\ln(\rho)+1}) & \tr(\tfrac{\rho}{2}\acomm{\mathbf{C_2}}{\mathbf{C_2}}) &\tr(\tfrac{\rho}{2}\acomm{\mathbf{C_2}}{\mathbf{C_3}})\\
					\tr(\tfrac{\rho}{2}\acomm{\mathbf{C_3}}{\ln(\rho)+1}) & \tr(\tfrac{\rho}{2}\acomm{\mathbf{C_3}}{\mathbf{C_2}}) & \tr(\tfrac{\rho}{2}\acomm{\mathbf{C_3}}{\mathbf{C_3}})},
			\end{split}\label{appBbeta1}                               \\
			\begin{split}
				\bar{\beta}_2 = & \dfrac{1}{k^2\tau^3\bar{\Omega}}\\
				&\times\vmqty{\tr(\tfrac{\rho}{2}\acomm{\mathbf{C_1}}{\ln(\rho)+1}) &\tr(\tfrac{\rho}{2}\acomm{\mathbf{C_1}}{\mathbf{C_1}}) &   \tr(\tfrac{\rho}{2}\acomm{\mathbf{C_3}}{\mathbf{C_1}}) \\
					\tr(\tfrac{\rho}{2}\acomm{\mathbf{C_2}}{\ln(\rho)+1}) &\tr(\tfrac{\rho}{2}\acomm{\mathbf{C_2}}{\mathbf{C_1}}) & \tr(\tfrac{\rho}{2}\acomm{\mathbf{C_2}}{\mathbf{C_3}})\\ \tr(\tfrac{\rho}{2}\acomm{\mathbf{C_3}}{\ln(\rho)+1}) &
					\tr(\tfrac{\rho}{2}\acomm{\mathbf{C_3}}{\mathbf{C_1}}) &  \tr(\tfrac{\rho}{2}\acomm{\mathbf{C_3}}{\mathbf{C_3}})},
			\end{split}\label{appBbeta2}                   \\
			\begin{split}
				\bar{\beta}_3 = & -\dfrac{1}{k^2\tau^3\bar{\Omega}}\\
				&\times\vmqty{\tr(\tfrac{\rho}{2}\acomm{\mathbf{C_1}}{\ln(\rho)+1}) &\tr(\tfrac{\rho}{2}\acomm{\mathbf{C_1}}{\mathbf{C_1}}) &  \tr(\tfrac{\rho}{2}\acomm{\mathbf{C_1}}{\mathbf{C_2}}) \\ \tr(\tfrac{\rho}{2}\acomm{\mathbf{C_2}}{\ln(\rho)+1})&
					\tr(\tfrac{\rho}{2}\acomm{\mathbf{C_2}}{\mathbf{C_1}}) & \tr(\tfrac{\rho}{2}\acomm{\mathbf{C_2}}{\mathbf{C_2}}) \\ \tr(\tfrac{\rho}{2}\acomm{\mathbf{C_3}}{\ln(\rho)+1})&
					\tr(\tfrac{\rho}{2}\acomm{\mathbf{C_3}}{\mathbf{C_1}}) & \tr(\tfrac{\rho}{2}\acomm{\mathbf{C_3}}{\mathbf{C_2}}) }.
			\end{split}\label{appBbeta3}
		\end{align}
	\end{minipage}}

% \endgroup
We consider equation (\ref{appaeD}) with $\mathcal{L}=\dfrac{1}{4k\tau}\mathrm{I}$, use the $\bar{\beta}_i$ from equations (\ref{appBomega}-\ref{appBbeta3}), and after some algebra involving determinants, arrive at the equation below,
\resizebox*{.8\linewidth}{!}{
	\begin{minipage}{\linewidth}
		\begin{align}\label{finalgetsuga}
			\begin{split}
				&\dv{\rho}{t} +i\comm{\mathcal{H}}{\rho} =\\
				&  -\dfrac{1}{\tau}\tfrac{\vmqty{\rho\ln(\rho) & \half\acomm{\mathbf{C_1}}{\rho} &  \half\acomm{\mathbf{C_2}}{\rho} &  \half \acomm{\mathbf{C_3}}{\rho}\\
						\tr(\tfrac{\rho}{2}\acomm{\mathbf{C_1}}{\ln(\rho)}) & \tr(\rho\mathbf{C_1}^2) &  \tr(\tfrac{\rho}{2}\acomm{\mathbf{C_1}}{\mathbf{C_2}}) & \tr(\tfrac{\rho}{2}\acomm{\mathbf{C_1}}{\mathbf{C_3}})\\ \tr(\tfrac{\rho}{2}\acomm{\mathbf{C_2}}{\ln(\rho)}) &\tr(\tfrac{\rho}{2}\acomm{\mathbf{C_2}}{\mathbf{C_1}}) & \tr(\rho\mathbf{C_2}^2) & \tr(\tfrac{\rho}{2}\acomm{\mathbf{C_1}}{\mathbf{C_3}})\\
						\tr(\tfrac{\rho}{2}\acomm{\mathbf{C_3}}{\ln(\rho)}) &	 \tr(\tfrac{\rho}{2}\acomm{\mathbf{C_3}}{\mathbf{C_1}}) & \tr(\tfrac{\rho}{2}\acomm{\mathbf{C_3}}{\mathbf{C_2}}) & \tr(\rho\mathbf{C_3}^2)}}{\vmqty{\tr(\tfrac{\rho}{2}\acomm{\mathbf{C_1}}{\mathbf{C_1}}) &  \tr(\tfrac{\rho}{2}\acomm{\mathbf{C_1}}{\mathbf{C_2}}) & \tr(\tfrac{\rho}{2}\acomm{\mathbf{C_1}}{\mathbf{C_3}})\\
						\tr(\tfrac{\rho}{2}\acomm{\mathbf{C_2}}{\mathbf{C_1}}) & \tr(\tfrac{\rho}{2}\acomm{\mathbf{C_2}}{\mathbf{C_2}}) &\tr(\tfrac{\rho}{2}\acomm{\mathbf{C_2}}{\mathbf{C_3}})\\
						\tr(\tfrac{\rho}{2}\acomm{\mathbf{C_3}}{\mathbf{C_1}}) & \tr(\tfrac{\rho}{2}\acomm{\mathbf{C_3}}{\mathbf{C_2}}) & \tr(\tfrac{\rho}{2}\acomm{\mathbf{C_3}}{\mathbf{C_3}})}}.
			\end{split}
		\end{align}
	\end{minipage}}

Thus we have a full-fledged equation of motion in $\rho$ under SEA. The $\beta_i$ as given in equations (\ref{appBomega}-\ref{appBbeta3}) can be rewritten in the following scaled form (using the scaling, $\bar{\Omega} = \frac{1}{(k\tau)^3}\Omega$),
\resizebox{0.9\linewidth}{!}{
	\begin{minipage}{\linewidth}
		\begin{align}
			\begin{split}
				\bar{\beta_1} = & -\dfrac{1}{k^2\tau^3\bar{\Omega}}\\
				&\times\vmqty{\tr(\tfrac{\rho}{2}\acomm{\mathbf{C_1}}{\ln(\rho)}) &  \tr(\tfrac{\rho}{2}\acomm{\mathbf{C_1}}{\mathbf{C_2}}) & \tr(\tfrac{\rho}{2}\acomm{\mathbf{C_1}}{\mathbf{C_3}})\\
					\tr(\tfrac{\rho}{2}\acomm{\mathbf{C_2}}{\ln(\rho)}) & \tr(\tfrac{\rho}{2}\acomm{\mathbf{C_2}}{\mathbf{C_2}}) &\tr(\tfrac{\rho}{2}\acomm{\mathbf{C_2}}{\mathbf{C_3}})\\
					\tr(\tfrac{\rho}{2}\acomm{\mathbf{C_3}}{\ln(\rho)}) & \tr(\tfrac{\rho}{2}\acomm{\mathbf{C_3}}{\mathbf{C_2}}) & \tr(\tfrac{\rho}{2}\acomm{\mathbf{C_3}}{\mathbf{C_3}})}, \\
				& = -k\beta_1
			\end{split}\label{appBbetaM1}             \\
			\begin{split}
				\bar{\beta_2} = & \dfrac{1}{k^2\tau^3\bar{\Omega}}\\
				&\times\vmqty{\tr(\tfrac{\rho}{2}\acomm{\mathbf{C_1}}{\ln(\rho)}) &\tr(\tfrac{\rho}{2}\acomm{\mathbf{C_1}}{\mathbf{C_1}}) &   \tr(\tfrac{\rho}{2}\acomm{\mathbf{C_3}}{\mathbf{C_1}}) \\
					\tr(\tfrac{\rho}{2}\acomm{\mathbf{C_2}}{\ln(\rho)}) &\tr(\tfrac{\rho}{2}\acomm{\mathbf{C_2}}{\mathbf{C_1}}) & \tr(\tfrac{\rho}{2}\acomm{\mathbf{C_2}}{\mathbf{C_3}})\\ \tr(\tfrac{\rho}{2}\acomm{\mathbf{C_3}}{\ln(\rho)}) &
					\tr(\tfrac{\rho}{2}\acomm{\mathbf{C_3}}{\mathbf{C_1}}) &  \tr(\tfrac{\rho}{2}\acomm{\mathbf{C_3}}{\mathbf{C_3}})}, \\
				& = k\beta_2
			\end{split}\label{appBbetaM2} \\
			\begin{split}
				\bar{\beta_3} = & -\dfrac{1}{k^2\tau^3\bar{\Omega}}\\
				&\times\vmqty{\tr(\tfrac{\rho}{2}\acomm{\mathbf{C_1}}{\ln(\rho)}) &\tr(\tfrac{\rho}{2}\acomm{\mathbf{C_1}}{\mathbf{C_1}}) &  \tr(\tfrac{\rho}{2}\acomm{\mathbf{C_1}}{\mathbf{C_2}}) \\ \tr(\tfrac{\rho}{2}\acomm{\mathbf{C_2}}{\ln(\rho)})&
					\tr(\tfrac{\rho}{2}\acomm{\mathbf{C_2}}{\mathbf{C_1}}) & \tr(\tfrac{\rho}{2}\acomm{\mathbf{C_2}}{\mathbf{C_2}}) \\ \tr(\tfrac{\rho}{2}\acomm{\mathbf{C_3}}{\ln(\rho)})&
					\tr(\tfrac{\rho}{2}\acomm{\mathbf{C_3}}{\mathbf{C_1}}) & \tr(\tfrac{\rho}{2}\acomm{\mathbf{C_3}}{\mathbf{C_2}}) }, \\
				& = -k\beta_3.
			\end{split}\label{appBbetaM3}
		\end{align}
	\end{minipage}}
This allows us to write equation (\ref{appaeD}) as follows,
\begin{equation}\label{sea_gen}
	\dv{\rho_D}{t} = -\dfrac{1}{\tau}\left[\rho\ln(\rho)+\half\sum_i(-1)^i\beta_i\acomm{\mathbf{C_i}}{\rho}\right].
\end{equation}
Single qubit will require only two $\mathbf{C_i}$s; hence, modifying the equation (\ref{finalgetsuga}) with appropriate constraints gets us the following equation
\begin{equation}\label{seaqa3}
	\dv{\rho}{t} +i\comm{\mathcal{H}}{\rho} =  -\dfrac{1}{\tau}\tfrac{\vmqty{\rho\ln(\rho) & \rho &  \half\acomm{\rho}{\mathcal{H}} \\
			\tr(\rho\ln(\rho)) & 1 &  \tr(\rho\mathcal{H}) \\ \tr(\rho\mathcal{H}\ln(\rho)) &		\tr(\rho\mathcal{H}) & \tr(\rho\mathcal{H}^2) \\
		}}{\vmqty{1 &  \tr(\rho\mathcal{H}) \\
			\tr(\rho\mathcal{H}) & \tr(\rho\mathcal{H}^2) }}.
\end{equation}
Considering qubits, as described in sub-section \ref{qubit} of the main text, we need to use the following traces as they appear in the equations for $\beta_i$ in equation (\ref{seaqa3}). We use $\mathcal{H}=\left(\omega_0\mathrm{I}+\omega\hat{h}\cdot\vec{\sigma}\right)$, and $\rho = \half\left(\mathrm{I}+\vec{r}\cdot\vec{\sigma}\right)$ as defined in the main text \cite{Beretta1985}. Defining $r_e = \hat{h}\cdot\vec{r}$ we get -
\begin{align*}
	\begin{split}
		\tr(\rho) =& 1,\\
		\tr(\rho\mathcal{H}) =& \left(\omega_0+\omega r_e\right),\\
		\tr(\rho\mathcal{H}^2) =& \left(\left(\omega^2+\omega_0^2\right)+2\omega_0\omega r_e\right),\\
		\tr(\rho\ln(\rho)) =& \half\left(\ln(\frac{1-r^2}{4})+r\ln(\frac{1+r}{1-r})\right),\\
		\tr(\rho\mathcal{H}\ln(\rho)) =& \dfrac{\omega_0}{2}\left(\ln(\frac{1-r^2}{4})+r\ln(\frac{1+r}{1-r})\right)\\
		&+ \dfrac{\omega}{2}\left(\ln(\frac{1-r^2}{4})+\frac{1}{r}\ln(\frac{1+r}{1-r})\right)r_e.
	\end{split}
\end{align*}

One can plug these traces into the $\beta$ expressions in equation (\ref{seaqa3}) above. Then along with the commutation/anticommutaion relations below, followed substituting all of these in equation (\ref{req}) to get the most general equation of motion for a single particle as shown below in equation (\ref{general}).
\begin{align*}
	\begin{split}
		\comm{\mathcal{H}}{\rho} = & i\omega\left(\hat{h}\cross\vec{r}\right)\cdot\vec{\sigma},\\
		\acomm{\mathcal{H}}{\rho} = & \left(\omega_0+\omega r_e\right)\mathrm{I}+\left(\omega_0\vec{r}+\omega\hat{h}\right)\cdot\vec{\sigma},\\
		\acomm{\ln(\rho)}{\rho} = & \half\left(\ln(\frac{1-r^2}{4})+r\ln(\frac{1+r}{1-r})\right)\mathrm{I}\\
		& + \half\left(\ln(\frac{1-r^2}{4})+\frac{1}{r}\ln(\frac{1+r}{1-r})\right)\vec{r}\cdot\vec{\sigma}.
	\end{split}
\end{align*}
Using the commutation mentioned above and the trace relations, we find the (unscaled) $\beta_i$'s to have the following expression (for the qubit case)
\begin{align}
	\beta_H & = \frac{k r_e}{2\omega\left(1-r_e^2\right)}\left[\mathrm{A}-\mathrm{B}\right], \label{seaqbetah}                                                                               \\
	\beta_I & = \dfrac{k}{2\omega^2\left(1-r_e^2\right)}\Bigl[\omega^2\left[\mathrm{A}- r_e^2\mathrm{B}\right] + \omega\omega_0 \left[\mathrm{A}-\mathrm{B}\right] \Bigr]. \label{seaqbetai}
\end{align}

Where, $\mathrm{A} = \left(\ln(\frac{1-r^2}{4})+r\ln(\frac{1+r}{1-r})\right)$, and $\mathrm{B}=\left(\ln(\frac{1-r^2}{4})+\frac{1}{r}\ln(\frac{1+r}{1-r})\right)$. The equation of motion in this case is, with $\dv{\rho}{t}= \half\dv{\vec{r}}{t}\cdot\vec{\sigma}$, and $r_e = 0$,
\begin{equation}\label{general}
	\dv{r}{t} = -\dfrac{1}{2\tau}\left(1-r^2\right)\ln(\dfrac{1+r}{1-r})
\end{equation}
The solution to the above equation is given in the main text as equation (\ref{berettaequation}). The form of equation (\ref{fullevolution}) and in extension equation (\ref{qubitevol}) is achieved by fixing the $\beta_i$'s using required conditions in equations (\ref{seaqbetah}), and (\ref{seaqbetai}). The above equations are strictly valid for $r<1$, and in extension $\abs{r_e}<1$.

%Some extra content in box####################################################################################################
%\left[\half\left((1-\omega^2)r_e-1\right)\mathrm{A}+\frac{r_e}{2}\left(\omega_0+\omega r_e\right)\mathrm{B}\right]\mathrm{I}#
%																															 #
%#############################################################################################################################																													
%------------------------------------------------------------------------------------------------------------------
\section{\label{Appendix3} Some comments on $\tau$}

$\tau$ appears as a relaxation time associated with the system itself. As discussed in the literature \cite{Beretta5,li2016von,li2018steepest,beretta2014steepest} it is associated with the speed of evolution of the state operator. Considering a Fisher-Rao metric, which for probability space turns into a uniform metric, one can write from equation (\ref{req11}) the following expression:
\begin{equation*}
	\dv{l}{t} = 2\sqrt{\dot{\gamma}_D\cdot\dot{\gamma}_D} = \dot{\epsilon}.
\end{equation*}
Here $\dot{\epsilon}$ is a small positive number, which fixes the norm of $\Pi_{\gamma}$ and maximizes the direction as a consequence \cite{beretta2014steepest}. From the evolution equation of state operator $\gamma_D$, we have,
\begin{equation*}
	\dot{\gamma}_D = \kae{\Pi_{\gamma}} = \frac{1}{\tau}\kae{\Phi-\sum_{i}\beta_i\mathbf{C_i}}.
\end{equation*}
Using these two and defining $\kae{\Lambda}$ as an affinity vector that draws the motion towards SEA evolution, one can write the following expressions involving $\tau$ \cite{beretta2014steepest}
\begin{align*}
	\tau & = \frac{\sqrt{\inprc{\Lambda}{\Lambda}}}{\dot{\epsilon}},                                                  \\
	     & = \frac{\seaexp{\Phi-\sum_{i}\beta_i\mathbf{C_i}}{\hat{G}^{-1}}{\Phi-\sum_{i}\beta_i\mathbf{C_i}}}{\Pi_S}.
\end{align*}
We see that $\tau$ is also inversely proportional to the entropy generation rate. As a result, for higher $\tau$, we will see lesser entropy generation and lesser dissipation; as the speed is high, the system doesn't relax quickly. On the other hand, the system will relax faster, and entropy generation will be enhanced in case of low $\tau$ values. Both of these features of $\tau$ are explored in the main text. As it can be seen, $\tau$ is dependent on $\rho$, yet as literature suggest, constant non-zero $\tau$ can also work in elucidating the features of the general motion.

%------------------------------------------------------------------------------------------------------------------
\section{\label{Appendix4}Computation of $\beta_i$'s for the quantum walker}
Considering, $ \rho_u = \dfrac{1}{N}\mathbf{I} $, and $ \mathcal{H} $ as given in equation (\ref{hamilctqw}), we begin by computing trace function as required in the equations (\ref{seaqa3}) which are as given below:
\begin{align*}
	\begin{split}
		\tr(\rho) =& 1,\\
		\tr(\rho\mathcal{H}) =& d,\\
		\tr(\rho\mathcal{H}^2) =&N\left(d^2+2\right),\\
		\tr(\rho\ln(\rho)) =& -\ln(N),\\
		\tr(\rho\mathcal{H}\ln(\rho)) =& -d\ln(N).
	\end{split}
\end{align*}
Using these traces and noticing that $\Omega$ is given by
\begin{equation*}
	\tr(\rho\mathcal{H}^2)\tr(\rho) - \left(\tr(\rho\mathcal{H})\right)^2 = \left(N-1\right)d^2+2N,
\end{equation*}
we can write the expressions below for $\ beta_i$s,
\begin{align}\label{ctqwbetas}
	\begin{split}
		\beta_H &= 0,\\
		\beta_I &= -\ln(N).\\
	\end{split}
\end{align}
We have suppressed the $ t$'s for notational convenience. This equation can use initial conditions and particular modeling parameters to give corresponding $\ beta_i$s.

%########################################
\bibliography{refer}

\end{document}